\documentclass[twoside]{article}
\usepackage{qic}
\usepackage{amsmath}

\newcommand{\uniset}{\mbox{$\mathcal{G}_{u}$}}
\newcommand{\capfx}{\mbox{$\Phi_{x}$}}
\newcommand{\capfz}{\mbox{$\Phi_{0}$}}
\newcommand{\tilfx}{\mbox{$\tilde{\Phi}_{x}$}}
\newcommand{\deltilfx}{\mbox{$\delta\tilde{\Phi}_{x}$}}
\newcommand{\tilfxz}{\mbox{$\tilde{\Phi}_{x}^{0}$}}
\newcommand{\calB}{\mbox{$\mathcal{B}$}}
\newcommand{\calF}{\mbox{$\mathcal{F}$}}
\newcommand{\calFn}{\mbox{$\mathcal{F}_{n}$}}
\newcommand{\bfF}{\mathbf{F}}
\newcommand{\bfB}{\mathbf{B}}
\newcommand{\bfs}{\mathbf{s}}
\newcommand{\hatbfx}{\hat{\mathbf{x}}}
\newcommand{\hatbfy}{\hat{\mathbf{y}}}
\newcommand{\hatbfz}{\hat{\mathbf{z}}}
\newcommand{\bfsig}{\mbox{\boldmath$\sigma$}}

\begin{document}
\setlength{\textheight}{8.0truein}    


\normalsize\textlineskip
\thispagestyle{empty}
\setcounter{page}{1}


\vspace*{0.88truein}

\alphfootnote

\fpage{1}

\centerline{\bf
HIGH FIDELITY UNIVERSAL SET OF QUANTUM GATES}
\vspace*{0.035truein}
\centerline{\bf USING NON-ADIABATIC RAPID PASSAGE}
\vspace*{0.37truein}
\centerline{\footnotesize
RAN LI}
\vspace*{0.015truein}
\centerline{\footnotesize\it Department of Physics, Kent State  
University, Stark Campus}
\baselineskip=10pt
\centerline{\footnotesize\it North Canton, OH 44720}
\vspace*{10pt}
\centerline{\footnotesize 
MELIQUE HOOVER}
\vspace*{0.015truein}
\centerline{\footnotesize\it Department of Physics, Southern Illinois 
University}
\baselineskip=10pt
\centerline{\footnotesize\it Carbondale, IL 62901-4401}
\vspace*{10pt}
\centerline{\footnotesize 
FRANK GAITAN\footnote{Corresponding author.}}
\vspace*{0.015truein}
\centerline{\footnotesize\it Department of Physics, Southern Illinois 
University}
\baselineskip=10pt
\centerline{\footnotesize\it Carbondale, IL 62901-4401}
\vspace*{0.225truein}

\vspace*{0.21truein}

\abstracts{
Numerical simulation results are presented which suggest that a class of
non-adiabatic rapid passage sweeps first realized experimentally
in 1991 should be capable of implementing a universal set of quantum gates
\uniset\ that operate with high fidelity. The gates constituting \uniset\ are
the Hadamard and NOT gates, together with variants of the phase, $\pi /8$, and 
controlled-phase gates. The universality of \uniset\ is established by showing 
that it can construct the universal set consisting of Hadamard, phase, 
$\pi /8$, and controlled-NOT gates. Sweep parameter values are provided which 
simulations indicate will produce the different gates in \uniset , and for
which the gate error probability $P_{e}$ satisfies: (i)~$P_{e}<10^{-4}$ for 
the one-qubit gates; and (ii)~$P_{e}<1.27\times 10^{-3}$ for the modified 
controlled-phase gate. The sweeps in this class are non-composite and generate 
controllable quantum interference effects that allow the gates in \uniset\ to 
operate non-adiabatically while maintaining high fidelity. These interference
effects have been observed using NMR, and it has previously been shown how 
these rapid passage sweeps can be applied to atomic systems using electric 
fields. Here we show how these sweeps can be applied to both superconducting 
charge and flux qubit systems. The simulations suggest that the universal set 
of gates \uniset\ produced by these rapid passage sweeps shows promise as 
possible elements of a fault-tolerant scheme for quantum computing.
}{}{}

\vspace*{10pt}

\vspace*{3pt}


\vspace*{1pt}\textlineskip    
\vspace*{-0.5pt}
\noindent
\section{\label{sec1}Introduction}        
Slightly more than a decade has passed since the accuracy threshold 
theorem was first proved, establishing the unexpected result that, under 
appropriate conditions, a quantum computation of arbitrary duration could be 
carried out with arbitrarily small error probability in the presence of noise, 
and using imperfect quantum gates \cite{ft1}--\cite{ft8}. The required 
conditions are that: (1)~the computational data is protected by a sufficiently 
layered concatenated quantum error correcting code; (2)~fault-tolerant 
protocols for quantum computation, error correction, and measurement are used; 
and (3)~all quantum gates used in the computation have error probabilities (per 
operation) $P_{e}$ that fall below a value known as the accuracy threshold 
$P_{a}$. This threshold has been calculated for a number of simple error 
models with results falling in the range $10^{-6}<P_{a}<10^{-3}$. For many, 
$P_{a}\sim 10^{-4}$ has become a rough-and-ready working estimate for the 
threshold so that gates are anticipated to be approaching the accuracies 
needed for fault-tolerant quantum computing when $P_{e}<10^{-4}$. A number of 
universal sets of quantum gates have been found \cite{us1}--\cite{boy}. With 
such a set of gates, any $N$-qubit unitary operation can be applied using a
quantum circuit made up solely of gates belonging to the universal set. Thus 
the problem of producing sufficiently accurate quantum gates shifts to 
producing a sufficiently accurate universal set of such gates. One well-known 
universal set consists of the one-qubit Hadamard, phase, and $\pi /8$ gates, 
together with the two-qubit controlled-NOT gate \cite{boy}. 

In this paper numerical simulation results are presented which suggest that a
class of non-adiabatic rapid passage sweeps known as twisted rapid passage
(TRP), first realized experimentally in 1991 \cite{zw1}, should be capable of 
implementing a universal set of quantum gates \uniset\ that operate 
non-adiabatically and with high fidelity. \uniset\ consists of the one-qubit 
Hadamard and NOT gates, together with variants of the: (i)~one-qubit phase and 
$\pi /8$ gates; and (ii)~two-qubit controlled-phase gate. The universality of 
this set is established by showing that it can construct the universal set of 
Ref.~\cite{boy}. Sweep parameter values are provided which simulations 
indicate will produce the different gates in \uniset , and for which the gate 
error probability $P_{e}$ satisfies: (i)~$P_{e}<10^{-4}$ for the one-qubit 
gates; and (ii)~$P_{e}<1.27\times 10^{-3}$ for the modified controlled-phase 
gate. This level of accuracy is a consequence of controllable quantum 
interference effects that are generated by this class of rapid passage sweeps 
\cite{trp1,trp2,lhg}. The sweep parameter values provided for each gate were 
found using an optimization procedure that searches for minima of $P_{e}$. 
TRP implementation of the one-qubit gates in \uniset\ was shown in 
Ref.~\cite{lhg}. Here we complete the universal set \uniset\ by showing that
TRP can also implement a high fidelity, non-adiabatic modified controlled-phase
gate. TRP sweeps have already been experimentally realized in NMR systems 
\cite{zw1,trp2}. Ref.~\cite{lhg} showed how TRP sweeps can be applied to atomic
systems using electric fields. Here we extend the list of physical systems to 
which TRP can be applied by showing how these sweeps can be applied to both 
superconducting charge and flux qubit systems.

The outline of this paper is as follows. Section~\ref{sec2} begins with a
summary of the essential properties of TRP. It then details our simulation 
protocol, and the procedure used to optimize the TRP sweep parameters. 
Section~\ref{sec3} presents our simulation results for the universal set of
gates \uniset . For each gate in \uniset , we present: the optimized sweep 
parameter values that produce it; an upper bound on the gate error probability 
$P_{e}$; and the gate fidelity $\mathcal{F}$. For the two-qubit modified 
controlled-phase gate, we also include the unitary operation produced. We have
not included the corresponding unitary operations for each of the one-qubit 
gates in \uniset\ as they appear in Ref.~\cite{lhg}. Section~\ref{sec4} 
then explains how TRP sweeps can be applied to both superconducting charge and 
flux qubits. Finally, Section~\ref{sec5} makes closing remarks.

\section{\label{sec2}Preliminaries}
\noindent This section discusses a number of important background topics. 
Section~\ref{sec2.1} provides a brief review of TRP and shows how these sweeps
generate quantum interference effects that are controllable through variation 
of the sweep parameters. Section~\ref{sec2.2} describes the protocol used to 
simulate the one- and two-qubit gates that are of interest in this paper. 
Finally, to produce high performance gates, the simulations are combined with 
an optimization procedure described in Section~\ref{sec2.3} that searches for 
sweep parameter values that minimize an upper bound on the gate error 
probability $P_{e}$. 

\subsection{\label{sec2.1}Twisted Rapid Passage}
\noindent To introduce TRP we consider a single qubit interacting with an 
external control field $\bfF (t)$ via the Zeeman interaction 
\begin{equation}
H_{Z}(t) = -\bfsig\cdot\bfF (t) , 
\label{zeeman}
\end{equation}
where $\bfsig_{i}$ are the Pauli matrices. The sweeps we will be interested in 
are a generalization of those used in adiabatic rapid passage (ARP) \cite{abr}. 
In ARP the field $\bfF (t)$ in the detector frame \cite{sut} is inverted over 
a time $T_{0}$ such that $\bfF (t)=b\,\hatbfx + at\,\hatbfz$. In the
NMR realization of ARP, as seen in the lab frame, the detector frame rotates
about the static magnetic field $B_{0}\,\hatbfz$. In the detector frame,
$\hatbfz$ is chosen parallel to the rotation axis. In the rotating wave
approximation the rf-magnetic field $\bfB_{rf}$ in the lab frame lies in the
$x$--$y$ plane and rotates about the static magnetic field. The detector frame
is chosen to rotate with $\bfB_{rf}$ so that, in this frame, the rf-field is
static and its direction defines $\hatbfx$: $\bfB_{rf} = b\,\hatbfx$. The
inversion time $T_{0}$ is large compared to the inverse Larmor frequency
$\omega_{0}^{-1}$ (viz.\ adiabatic), though small compared to the thermal
relaxation time $\tau_{th}$ (viz.\ rapid). ARP sweeps provide a highly precise
method for inverting the Bloch vector $\bfs_{i}=\langle\bfsig_{i}\rangle$,
although the price paid for this precision is an adiabatic inversion rate.
We are interested in a type of rapid passage in which the control field
$\bfF (t)$, as seen in the detector frame, is allowed to twist around in the
$x$--$y$ plane with time-varying azimuthal angle $\phi (t)$, while 
simultaneously undergoing inversion along the $z$-axis:
\begin{equation}
\bfF (t) = b\,\cos\phi (t)\,\hatbfx +b\,\sin\phi (t)\,\hatbfy +at\,\hatbfz
 \hspace{0.1in} .
\label{TRPsweep}
\end{equation}
Here $-T_{0}/2\leq t\leq T_{0}/2$ and $\hatbfy = \hatbfz\times\hatbfx$. Note
that any pair of orthogonal unit vectors in the $x$--$y$ plane can be used for
$\hatbfx$ and $\hatbfy$. Different choices simply alter the value of $\phi 
(t=0)$. As will be seen shortly, interesting physical effects arise when the
twist profile $\phi (t)$ is chosen appropriately. This class of rapid passage
sweeps is referred to as twisted rapid passage (TRP). The first experimental
realization of TRP in 1991 by Zwanziger et al.\ \cite{zw1} carried out the
inversion adiabatically with $\phi (t)=Bt^{2}$. Subsequently, 
\textit{non-adiabatic\/} TRP was studied with polynomial twist profile 
$\phi (t)=(2/n)Bt^{n}$ \cite{trp1}, and controllable quantum interference 
effects were found to arise for $n\geq 3$. Zwanziger et al.\ \cite{trp2} 
implemented non-adiabatic polynomial TRP with $n=3,4$ and observed the 
predicted interference effects. In the Zwanziger experiments \cite{zw1,trp2}, 
a TRP sweep is produced by sweeping the detector frequency $\dot{\phi}_{det}
(t)$ linearly through resonance at the Larmor frequency $\omega_{0}$: 
$\dot{\phi}_{det}(t)=\omega_{0}+(2at)/\hbar$. The frequency of the rf-field
$\dot{\phi}_{rf}(t)$ is also swept through resonance in such a way that
$\dot{\phi}_{rf}(t)=\dot{\phi}_{det}(t)-\dot{\phi}(t)$, where $\phi (t)=(2/n)
Bt^{n}$ is the TRP polynomial twist profile. Substituting the expression for 
$\dot{\phi}_{det}(t)$ into that for $\dot{\phi}_{rf}(t)$ gives
\begin{equation}
\dot{\phi}_{rf}(t) = \omega_{0} +\frac{2at}{\hbar} - \dot{\phi}(t)
 \hspace{0.1in} .
\label{phirf}
\end{equation}
At resonance $\dot{\phi}_{rf}(t)=\omega_{0}$. Inserting this condition into
eq.~(\ref{phirf}), it follows that at resonance
\begin{equation}
at -\frac{\hbar}{2}\dot{\phi}(t) = 0 \hspace{0.1in} .
\label{rescond}
\end{equation}
For polynomial twist $\phi (t)=(2/n)Bt^{n}$, eq.~(\ref{rescond}) has $n-1$
roots, though only real-valued roots correspond to resonance. Ref.~\cite{trp1}
showed that for $n\geq 3$, multiple passes through resonance occur during a
\textit{single\/} TRP sweep: (i)~for all $n$ when $B>0$; and (ii)~for $n$ odd
when $B<0$. We restrict ourselves to $B>0$ in the remainder of this paper. In 
this case, the qubit passes through resonance at the times:
\begin{equation}
t = \left\{\hspace{0.1in} \begin{array}{ll}
                             0,\;  \left( a/\hbar B\right)^{\frac{1}{n-2}}
                                  & (n\;\;\mathrm{odd})\\
                             0,\; \pm\left( a/\hbar B\right)^{\frac{1}{n-2}}
                                  & (n\;\;\mathrm{even})\\
                          \end{array} 
    \right. \hspace{0.1in} .
\label{restimes}
\end{equation}
We see that the time separating the qubit resonances can be altered by 
variation of the sweep parameters $B$ and $a$. Ref.~\cite{trp1} showed that
these multiple resonances have a strong influence on the qubit transition
probability. It was shown that qubit transitions could be significantly
enhanced or suppressed by small variation of the sweep parameters, and
hence of the time separating the resonances. Plots of the transition
probability versus time suggested that the multiple resonances were producing
quantum interference effects that could be controlled by variation of the
TRP sweep parameters. In Ref.~\cite{fg} the qubit transition amplitude was
calculated to all orders in the non-adiabatic coupling. The result found there
can be re-expressed as the following diagrammatic series:
\begin{equation}
\label{diagseries}
\setlength{\unitlength}{0.05in}
T_{-}(t) = \begin{picture}(10,5)
              \put(10,-1.5){\vector(-1,0){3.25}}
              \put(5,-1.5){\line(1,0){1.75}}
              \put(5,-1.5){\vector(0,1){3.25}}
              \put(5,1.75){\line(0,1){1.75}}
              \put(5,3.5){\vector(-1,0){3.25}}
              \put(0,3.5){\line(1,0){1.75}}
           \end{picture}
\hspace{0.05in} +
           \begin{picture}(20,5)
              \put(20,-1.5){\vector(-1,0){3.25}}
              \put(15,-1.5){\line(1,0){1.75}}
              \put(15,-1.5){\vector(0,1){3.25}}
              \put(15,1.75){\line(0,1){1.75}}
              \put(15,3.5){\vector(-1,0){3.25}}
              \put(10,3.5){\line(1,0){1.75}}
              \put(10,3.5){\vector(0,-1){3.25}}
              \put(10,-1.5){\line(0,1){1.75}}
              \put(10,-1.5){\vector(-1,0){3.25}}
              \put(5,-1.5){\line(1,0){1.75}}
              \put(5,-1.5){\vector(0,1){3.25}}
              \put(5,1.75){\line(0,1){1.75}}
              \put(5,3.5){\vector(-1,0){3.25}}
              \put(0,3.5){\line(1,0){1.75}}
           \end{picture}
\hspace{0.05in} +
           \begin{picture}(30,5)
              \put(30,-1.5){\vector(-1,0){3.25}}
              \put(25,-1.5){\line(1,0){1.75}}
              \put(25,-1.5){\vector(0,1){3.25}}
              \put(25,1.75){\line(0,1){1.75}}
              \put(25,3.5){\vector(-1,0){3.25}}
              \put(20,3.5){\line(1,0){1.75}}
              \put(20,-1.5){\vector(-1,0){3.25}}
              \put(20,3.5){\vector(0,-1){3.25}}
              \put(20,-1.5){\line(0,1){1.75}}
              \put(15,-1.5){\line(1,0){1.75}}
              \put(15,-1.5){\vector(0,1){3.25}}
              \put(15,1.75){\line(0,1){1.75}}
              \put(15,3.5){\vector(-1,0){3.25}}
              \put(10,3.5){\line(1,0){1.75}}
              \put(10,3.5){\vector(0,-1){3.25}}
              \put(10,-1.5){\line(0,1){1.75}}
              \put(10,-1.5){\vector(-1,0){3.25}}
              \put(5,-1.5){\line(1,0){1.75}}
              \put(5,-1.5){\vector(0,1){3.25}}
              \put(5,1.75){\line(0,1){1.75}}
              \put(5,3.5){\vector(-1,0){3.25}}
              \put(0,3.5){\line(1,0){1.75}}
           \end{picture}
\hspace{0.05in} + \hspace{0.05in} \cdots \hspace{0.25in} .
\end{equation}
Lower (upper) lines correspond to propagation in the negative (positive)
energy level and the vertical lines correspond to transitions between the
two energy levels. The calculation sums the probability amplitudes for all
interfering alternatives \cite{fh} that allow the qubit to end up in the
positive energy level at time $t$ given that it was initially in the negative
energy level. As we have seen, varying the TRP sweep parameters varies the
time separating the resonances. This in turn changes the value of each
diagram in eq.~(\ref{diagseries}), and thus alters the interference between
alternatives in the quantum superposition. Similar diagrammatic series can be
worked out for the remaining $3$ combinations of final and initial states.
It is the sensitivity of the individual alternatives/diagrams to the time
separation of the resonances that allow TRP to manipulate this quantum
interference. Zwanziger et al.\ \cite{trp2} observed these interference effects
in the transition probability using liquid state NMR and found quantitative
agreement between theory and experiment. It is the link between the TRP
sweep parameters and this quantum interference that we believe makes it
possible for TRP to drive highly accurate non-adiabatic one- and two-qubit 
gates. The results presented in Section~\ref{sec3} for the different gates
in the universal set \uniset\ are found by numerical simulation of the
Schrodinger equation. We next describe how the simulations are done.

\subsection{\label{sec2.2}Simulation Protocol}
\noindent As is well-known, an $N$-qubit quantum gate applies a fixed unitary 
transformation $U$ to $N$-qubit states:
\begin{equation}
|\psi_{out}\rangle = U\, |\psi_{in}\rangle . 
\end{equation}
We will be interested in the unitary transformations applied by the: 
(i)~one-qubit Hadamard ($U_{H}$), phase ($U_{P}$), $\pi /8$ ($U_{\pi /8}$), 
and NOT ($U_{\scriptscriptstyle NOT}$) gates 
\begin{equation}
\hspace{-0.6in} U_{H} = \frac{1}{\sqrt{2}}\left( \begin{array}{cc}
                                    1 & 1 \\
                                    1 & -1
                                 \end{array}
                          \right)
 \hspace{0.40in} ; \hspace{0.15in} 
U_{P} = \left( \begin{array}{cc}
                  1 & 0 \\
                  0 & i 
               \end{array}
        \right) \vspace{0.1in} 
\label{Udefs1}
\end{equation}
\begin{equation}
 \hspace{-0.45in} U_{\pi /8} = \left( \begin{array}{cc}
                       1 & 0 \\
                       0 & e^{i\pi /4}
                    \end{array}
             \right)
 \hspace{0.45in} ; \hspace{0.15in} 
U_{\scriptscriptstyle NOT} = \left( \begin{array}{cc}
                                       0 & 1 \\
                                       1 & 0 
                                    \end{array}
                             \right) , \vspace{0.10in} 
\label{Udefs2}
\end{equation}
and (ii)~the two-qubit controlled-NOT ($U_{\scriptscriptstyle CNOT}$) and 
controlled-phase ($U_{\scriptscriptstyle CP}$) gates
\begin{equation}
U_{\scriptscriptstyle CNOT} = \left( \begin{array}{cccc}
                                         1 & 0 & 0 & 0 \\
                                         0 & 1 & 0 & 0 \\
                                         0 & 0 & 0 & 1 \\
                                         0 & 0 & 1 & 0
                                      \end{array}
                               \right) \hspace{0.15in}  
  ; \hspace{0.15in} 
U_{\scriptscriptstyle CP} = \left( \begin{array}{cccc}
                                      1 & 0 & 0 & 0 \\
                                      0 & 1 & 0 & 0 \\
                                      0 & 0 & 1 & 0 \\
                                      0 & 0 & 0 & -1
                                   \end{array}
                            \right) .
\label{Udefs3}
\end{equation}
The one- and two-qubit matrices appearing in 
eqs.~(\ref{Udefs1})--(\ref{Udefs3}) are in the representation spanned by the 
one- and two-qubit computational basis states $|i\rangle$ and $|ij\rangle$ 
which are, respectively, the eigenstates of $\sigma_{z}$ and $\sigma_{z}^{1}
\otimes\sigma_{z}^{2}$:
\begin{equation}
\sigma_{z}\, |i\rangle = \left( -1\right)^{i}|i\rangle \hspace{0.25in} ;
  \hspace{0.25in}
 \sigma_{z}^{1}\otimes\sigma_{z}^{2}\, |ij\rangle = \left( -1\right)^{i+j}
  |ij\rangle \hspace{0.25in} (i,j=0,1) .
\end{equation}

The dynamical impact of TRP is determined by numerical simulation of the 
Schrodinger equation. Let $H(t)$ denote the Hamiltonian for an $N$-qubit 
system, and $|E_{k}(t)\rangle$ the instantaneous energy eigenstates which 
satisfy $H(t)\,
|E_{k}(t)\rangle = E_{k}(t)\, |E_{k}(t)\rangle$ with $k=1,\ldots , 2^{N}$. It 
is found that the numerical stability of the simulations is enhanced if we 
expand $|\psi (t)\rangle$ in the instantaneous energy eigenstates $|E_{k}(t)
\rangle$. Because of the direct connection between these states and $H(t)$, 
they carry substantial dynamical information, and a substantial portion of the 
dynamics due to $H(t)$ can be accounted for by choosing  this basis. This 
makes the task of determining the remaining dynamics using the Schrodinger 
equation much simpler and the simulations more stable. We thus write
\begin{equation}
|\psi (t)\rangle = \sum_{k=1}^{2^{N}}\, a_{k}(t)\, |E_{k}(t)\rangle\,
                    \exp\left[ -\frac{i}{\hbar}\int_{-T_{0}/2}^{t}\,
                     d\tau\left\{ \, E_{k}(\tau )-\hbar\dot{\gamma}_{k}(\tau )
                      \right\} \right] ,
\label{stdecomp}
\end{equation}
where $\gamma_{k}(t)$ is the adiabatic geometric phase \cite{sh+w} associated
with the energy level $E_{k}(t)$, and 
\begin{equation}
\dot{\gamma}_{k}(t) = i\,\langle E_{k}(t)|\frac{d}{dt}|E_{k}(t)\rangle .
\end{equation}
Substituting eq.~(\ref{stdecomp}) into the Schrodinger equation, and using the
orthonormality of the instantaneous energy eigenstates, one arrives at the 
equations of motion for the expansion coefficients $a_{k}(t)$:
\begin{equation}
\frac{da_{k}}{dt} = -\sum_{l\neq k}\, a_{l}(t)\,\Gamma_{kl}(t)\,
                      \exp\left[ -i\int_{-T_{0}/2}^{t} d\tau\:
                       \delta_{lk}(\tau )\right] ,
\label{eqsmotn}
\end{equation}
where $k=1,\ldots ,2^{N}$, and
\begin{eqnarray}
\Gamma_{kl}(t) & = & \langle E_{k}(t)|\frac{d}{dt}|E_{l}(t)\rangle 
   \label{Gammadef}\\
\delta_{lk}(t) & = & \frac{(E_{l}(t)-E_{k}(t))}{\hbar} -\left(
                      \dot{\gamma}_{l}(t) -\dot{\gamma}_{k}(t)\right) .
\label{deltadef}
\end{eqnarray}
The simulations set the initial $N$-qubit state to be one of the computational 
basis states $|i_{1}\cdots i_{2^N}\rangle$. The simulation outcome is the 
$|i_{1}\cdots i_{2^N}\rangle$-column of the unitary transformation $U$ 
produced by the TRP sweep. By simulating all $2^{N}$ computational basis 
states, $U$ is determined column-by-column.  

It proves useful to recast eqs.~(\ref{eqsmotn}) in dimensionless form.
For the one-qubit case we introduce the dimensionless time $\tau =(a/b)t$,
the dimensionless inversion rate $\lambda = \hbar a/b^{2}$, and the
dimensionless twist strength $\eta_{n}=(\hbar B/a)(b/a)^{n-2}$. For the
remainder of this paper we restrict ourselves to quartic TRP which has twist
profile $\phi_{n}(\tau )$ with $n=4$:
\begin{equation}
\phi_{4}(\tau ) = \frac{1}{2}B\left(\frac{b\tau}{a}\right)^{4}=
                   \left( \frac{\eta_{\scriptscriptstyle 4}}{2\lambda}\right) 
                    \tau^{4} .
\label{quarticTRP}
\end{equation}
The sweep parameters $a$, $b$, and $B$ are chosen to be positive so that
$\lambda$ and $\eta_{\scriptscriptstyle 4}$ are also positive. The Hamiltonian
for the one-qubit gates is the Zeeman Hamiltonian $H_{Z}(t)$ given in 
eqs.~(\ref{zeeman}) and (\ref{TRPsweep}). The dimensionless version of
$H_{Z}(t)$ is found by multiplying it by $(b/\hbar a)$. Denoting the result
by $H_{1}(\tau )$, one finds
\begin{equation}
H_{1}(\tau ) = -\frac{\tau}{\lambda}\,\sigma_{z} - \frac{1}{\lambda}\left[\,
                  \cos\phi_{4}\,\sigma_{x} +\sin\phi_{4}\,\sigma_{y}\,\right] 
                    , 
\end{equation}
where $\phi_{4}(\tau )$ is given by eq.~(\ref{quarticTRP}). $H_{1}(\tau )$
is the Hamiltonian that drives the one-qubit simulations via 
eqs.~(\ref{eqsmotn})--(\ref{deltadef}) in dimensionless form.

The derivation of the dimensionless Hamiltonian $H_{2}(\tau )$ that drives
the two-qubit simulations is more complicated and is presented in Appendix~A.
As explained there, $H_{2}(\tau )$ includes an interaction that Zeeman-couples
each qubit to a TRP control field, as well as an Ising interaction between
the two qubits. The Ising interaction was chosen because of its simplicity, 
and because of its occurrence in many physical systems. Modification of the
simulations to include alternative qubit-qubit interactions is straightforward.
The resulting two-qubit Hamiltonian yields energy level spacings that give 
rise to a degeneracy in the resonance energy for the energy level pairs 
($E_{1}\leftrightarrow E_{2}$) and ($E_{3}\leftrightarrow E_{4}$). To remove 
this degeneracy the term $\Delta H = c_{4}\, |E_{4}(t)\rangle\langle E_{4}
(t)|$ is added to the Hamiltonian, where $c_{4}$ is a constant. The end result 
is the Hamiltonian $H_{2}(\tau )$ that drives the two-qubit simulations:
\begin{eqnarray}
H_{2}(\tau ) & = & 
                   \left[\, -\frac{(d_{1}+d_{2})}{2} + \frac{\tau}{\lambda}
                    \,\right]\,\sigma_{1z} -\frac{d_{3}}{\lambda}
                     \left[\,\cos\phi_{4}\,\sigma_{1x}+\sin\phi_{4}\,
                      \sigma_{1y}\,\right] \nonumber\\
  & & \hspace{0.15in}
          +        \left[\, -\frac{d_{2}}{2} + \frac{\tau}{\lambda}
                    \,\right]\,\sigma_{2z} -\frac{1}{\lambda}
                     \left[\,\cos\phi_{4}\,\sigma_{2x}+\sin\phi_{4}\,
                      \sigma_{2y}\,\right] \nonumber\\
 & & \hspace{0.3in} -\frac{\pi}{2}d_{4}\,\sigma_{1z}\sigma_{2z} + 
            \Delta H .
\label{simHam}
\end{eqnarray}
The constants $d_{i}$ ($i=1,\ldots ,4$) are defined in Appendix~A where a
description of their physical significance is also given. 

Having described how the simulations are done, and how we determine the actual 
unitary transformation $U_{a}$ produced by a specific assignment of the TRP 
sweep parameters, we go on in Section~\ref{sec2.3} to explain how the 
sweep parameters are iteratively modified so as to make $U_{a}$ approach a 
target gate $U_{t}$ as closely as possible.

\subsection{\label{sec2.3}Sweep Parameter Optimization}
\noindent Let $\mathcal{H}_{N}$ be the Hilbert space for an $N$-qubit system.
As in Section~\ref{sec2.2}, let $U_{a}$ denote the actual unitary operation 
produced by a given set of TRP sweep parameters, and $U_{t}$ a target unitary 
operation we would like TRP to approximate as closely as possible. 
Introducing the operators $D=U_{a}-U_{t}$ and $P=D^{\dagger}D$, and the 
normalized state $|\psi\rangle$, we define $|\psi_{a}\rangle = U_{a}|\psi
\rangle$, $|\psi_{t}\rangle = U_{t}|\psi\rangle$, and $|\psi_{\perp}\rangle =
\{\, I -|\psi_{t}\rangle\langle\psi_{t}|\, \}|\psi_{a}\rangle$. Ref.~\cite{lhg}
showed that the error probability $P_{e}(\psi )$ for $U_{a}$ acting on the
state $|\psi\rangle$ is $P_{e}(\psi ) = |\langle\psi_{\perp}|\psi_{\perp}
\rangle |^{2}$. The gate error probability $P_{e}$ for the TRP gate $U_{a}$ was
then defined to be the worst-case error probability:
\begin{equation}
P_{e} \equiv \max_{\scriptstyle |\psi\rangle}\, P_{e}(\psi ) .
\label{Pedef}
\end{equation}
Because of the search over $|\psi\rangle$, evaluation of $P_{e}$ from 
eq.~(\ref{Pedef}) is not practical. Ref.~\cite{lhg} showed that
\begin{equation}
P_{e}\leq Tr\, P , 
\end{equation}
where $P=D^{\dagger}D$ was defined above. Once $U_{a}$ has been determined by
the numerical simulation, $Tr\, P$ is easily calculated, and so it makes a 
convenient proxy for $P_{e}$ in the sweep parameter optimization procedure 
that we are now ready to describe.

To find TRP sweep parameter values that yield highly accurate non-adiabatic
quantum gates it proved necessary to combine our simulations with function
minimization algorithms \cite{numrec} that search for sweep parameters that 
minimize the upper bound $Tr\, P$ for the gate error probability $P_{e}$.
The multi-dimensional downhill simplex method was used for the one-qubit gates
in the universal set \uniset , while simulated annealing was used for the
two-qubit modified controlled-phase gate. For quartic TRP, the sweep parameters
are ($\lambda$,$\:\eta_{\scriptscriptstyle 4}$) which can be thought of as 
specifying a point in
a two-dimensional parameter space. The downhill simplex method takes as input
three sets of sweep parameters which specify the vertices of a simplex in the
two-dimensional parameter space. The dynamical effect of the TRP sweep
associated with each vertex is found by numerically integrating the Schrodinger
equation as described in Section~\ref{sec2.2}. The output of the integration
is the unitary operation $U_{a}$ that a particular sweep applies. Having 
$U_{a}$ we determine $P=(U_{a}-U_{t})^{\dagger}(U_{a}-U_{t})$ and evaluate 
$Tr\, P$. The downhill simplex method iteratively alters the simplex (i.e.\ 
one or more of its vertices) until sweep parameter values are found that yield
a local minimum of $Tr\, P$. Because the minimum found is not global, some 
starting simplexes will give deeper minima than others. Though there was no 
guarantee, it was hoped that a starting simplex could be found that yielded 
$Tr\, P<10^{-4}$. Some trial and error in specifying the starting simplex was 
thus required, though for the one-qubit gates in \uniset , the trial and error
procedure eventually proved successful. For the two-qubit modified 
controlled-phase gate, trial and error proved inadequate, and it was necessary
to use simulated annealing to find sweep parameter values that yielded $Tr\, 
P\sim 10^{-3}$. We present our results for all gates in \uniset\ in the 
following section.

\section{\label{sec3}Universal Set of Quantum Gates}
\noindent To determine fidelities for the TRP gates we use the fidelity
\begin{equation}
\calFn = \frac{1}{2^{n}}\,\mathrm{Re}\left[ Tr\left(\, U_{a}^{\dagger}
           U_{t}\,\right)\right] ,
\label{fiddef}
\end{equation}
where $n$ denotes the number of qubits acted on by the gate. \calFn\ 
is an extension to $n$ qubits of the fidelity used in Ref.~\cite{mort} 
for one-qubit gates ($n=1$). It is possible to relate our $Tr\, P$ upper bound 
for $P_{e}$ to \calFn . Recalling that $P=(U_{a}-U_{t})^{\dagger}\, (U_{a}-
U_{t})$, it follows that
\begin{eqnarray}
Tr\, P & = & Tr\,\left[\, 2 -\left( U_{a}^{\dagger}U_{t}+U_{t}^{\dagger}U_{a}
                   \right)\right] \nonumber\\
 & = & 2^{n+1} - 2\,\mathrm{Re}\left[\, Tr\,\left( U_{a}^{\dagger}U_{t}\right)
                               \right]\nonumber\\
 & = & 2^{n+1}\left( 1 -\calFn\right) ,
\end{eqnarray}
and so 
\begin{equation}
\calFn = 1 - \left(\frac{1}{2^{n+1}}\right)\, Tr\, P .
\label{fidtrplink}
\end{equation}
$Tr\, P$ thus yields the fidelity \calFn\ and an upper
bound on the gate error probability $P_{e}$.

Next, notice that the gates $U_{P}$ and $U_{\pi /8}$ (see 
eqs.~(\ref{Udefs1}) and (\ref{Udefs2})) can be re-written as
\begin{eqnarray}
U_{P} & = & e^{i\pi /4}\, U_{\scriptscriptstyle NOT}\, V_{P}
   \label{uniproof1}\\
U_{\pi /8} & = & e^{i\pi /8}\, U_{\scriptscriptstyle NOT}\, V_{\pi /8},
\label{uniproof2}
\end{eqnarray}
where
\begin{eqnarray}
V_{P} & = & \left( \begin{array}{cc}
                      0 & e^{i\pi /4}\\
                      e^{-i\pi /4} & 0
                   \end{array}
            \right) \label{Vpdef}\\
V_{\pi /8} & = & \left( \begin{array}{cc}
                      0 & e^{i\pi /8}\\
                      e^{-i\pi /8} & 0
                   \end{array}
            \right) , \label{Vp8def}
\end{eqnarray}
and $U_{\scriptscriptstyle NOT}$ is given in eq.~(\ref{Udefs2}). Note also 
that $U_{\scriptscriptstyle CNOT}$ (eq.~(\ref{Udefs3})) can be re-written as
\begin{equation}
U_{\scriptscriptstyle CNOT} =\left( I^{1}\otimes U_{H}^{2}\right)
               \left[\,\left(\,\sigma_{z}^{1}\otimes I^{2}\right)\, 
                 V_{\scriptscriptstyle CP}\,\right]\,\left( I^{1}\otimes
                  U_{H}^{2}\right) .
\label{uniproof3}
\end{equation} 
Here the superscript on a one-qubit gate labels the qubit on which the gate
acts, and we have introduced the modified controlled-phase gate
\begin{equation}
V_{\scriptscriptstyle CP} = \left( \begin{array}{cccc}
                                     1 & 0 & 0 & 0 \\
                                     0 & 1 & 0 & 0 \\
                                     0 & 0 & -1 & 0 \\
                                     0 & 0 & 0 & 1 \\
                                   \end{array}
                            \right) .
\label{modcp}
\end{equation}
Finally, notice that the Pauli matrix $\sigma_{z}^{i}$ can be implemented 
using the phase gate $U_{P}^{i}$: $\sigma_{z}^{i} = \left( U_{P}^{i}
\right)^{2}$. It follows from this and eqs.~(\ref{uniproof1}), 
(\ref{uniproof2}), and (\ref{uniproof3}) that the set of gates $\{ U_{H}$,
$U_{P}$, $U_{\pi /8}$, $U_{\scriptscriptstyle CNOT}\,\}$ can be constructed 
using the set $\uniset = \left\{ U_{H},U_{\scriptscriptstyle NOT},V_{P},
V_{\pi /8},V_{\scriptscriptstyle CP}\right\}$. Since the first set of gates is
universal \cite{boy}, so is the set \uniset . As will be seen below, TRP can
be used to produce all gates in \uniset . For each gate we present our 
best-case result and show how gate performance is altered by small variation 
of the parameters.

\subsection{\label{sec3.1}One-Qubit Gates}
\noindent A study of the TRP-implementation of the one-qubit gates in 
\uniset\ was first reported in Ref.~\cite{lhg}. The essential results are 
included here for the reader's convenience, though for the sake of brevity, we 
have not reproduced the unitary operation $U_{a}$ generated by TRP for each 
gate. The interested reader can find them displayed in Ref.~\cite{lhg}.

As noted in Section~\ref{sec2.2}, all results presented in this paper are for 
quartic TRP which has twist profile
\begin{equation}
\phi (\tau ) = \frac{1}{2}\left(\frac{\eta_{\scriptscriptstyle 4}}{\lambda}
\right)\tau^{4}.
\end{equation}
Here $\tau$, $\lambda$, and $\eta_{\scriptscriptstyle 4}$ are dimensionless 
versions of the time $t$, inversion rate $a$, and twist strength $B$. For the 
reader's 
convenience, their definitions (Section~\ref{sec2.1}) are restated here:
\begin{equation}
\tau = \left(\frac{a}{b}\right) t \hspace{0.25in} ; \hspace{0.25in}
 \lambda = \frac{\hbar a}{b^{2}} \hspace{0.25in} ; \hspace{0.25in}
  \eta_{\scriptscriptstyle 4} = \left(\frac{\hbar b^{2}}{a^{3}}\right) B .
\label{paramdefs}
\end{equation}
Throughout this paper we assume $a$, $b$, and $B$ are positive. The parameter 
$b$ was introduced in eq.~(\ref{TRPsweep}) and is proportional to the 
rf-field amplitude in an NMR realization of TRP \cite{trp1,trp2}. All 
simulations were done with $\lambda >1$ corresponding to non-adiabatic 
inversion \cite{trp1,trp2}, and the dimensionless inversion time $\tau_{0}=a
T_{0}/b$ was fixed at $80.000$ for the one-qubit simulations, and at $120.00$ 
for the two-qubit simulations. 

The translation key connecting the simulation parameters to the experimental
sweep parameters used in the (one-qubit) Zwanziger experiments \cite{zw1,trp2} 
is given in the Appendix of Ref.~\cite{trp1}. We re-write the formulas for 
quartic twist here for convenience. Note that Zwanziger's symbol $B$ is here 
replaced by \calB\ to avoid confusion with our use of $B$ to denote 
the twist strength. First we give the formulas connecting our parameters
($a$,$\, b$,$\, B$,$\, T_{0}$) to the Zwanziger parameters ($\omega_{1}$,$\,
A$,$\,\calB ,T_{0}$):
\begin{eqnarray}
\omega_{1} & = & \frac{2b}{\hbar}\\
A & = & \frac{aT_{0}}{\hbar}\\
\calB & = & \frac{BT_{0}^{4}}{2} ,
\end{eqnarray}
where the inversion time $T_{0}$ is common to both parameter sets.
The formulas linking the dimensionless sweep parameters 
($\lambda$,$\,\eta_{\scriptscriptstyle 4}$) to the Zwanziger parameters 
($\omega_{1}$,$\, A$, $\,\calB$,$\, T_{0}$) are:
\begin{eqnarray}
\lambda & = & \frac{4A}{\omega_{1}^{2}T_{0}} \\
\eta_{\scriptscriptstyle 4} & = & \frac{\calB\omega_{1}^{2}}{2A^{3}T_{0}} .
\label{expeta}
\end{eqnarray}
In the experiments of Ref.~\cite{trp2}: $\omega_{1}=393\, \mathrm{Hz}$; 
$T_{0}=41.00\, \mathrm{ms}$; $A=50\, 000\, \mathrm{Hz}$; and \calB\ was 
calculated from eq.~(\ref{expeta}) with $\eta_{\scriptscriptstyle 4}$ varying 
over the range $\mathrm{[}4.50,4.70 \mathrm{]}\times 10^{-4}$.

\subsubsection*{\underline{Hadamard Gate}}
\noindent Here the target gate $U_{t}$ is the Hadamard gate $U_{H}$ (see
eq.~(\ref{Udefs1})). The TRP sweep parameters $\lambda = 5.8511$ and 
$\eta_{\scriptscriptstyle 4}= 2.9280\times 10^{-4}$ produce a unitary gate 
$U_{a}$ \cite{lhg} for which $Tr\, P = 8.82\times 10^{-6}$. This yields a gate 
fidelity $\calF_{H}=0.9999\; 98$ and the bound $P_{e}\leq 8.82 \times 10^{-6}$ 
on the gate error probability. Table~\ref{table1} shows how gate performance 
varies when the sweep parameters are altered slightly.
\begin{table}[ht!]
\begin{center}
\tcaption{\label{table1}Variation of $Tr\, P$ for the Hadamard gate when the
TRP sweep parameters are altered slightly from their best performance values.
The columns to the left of center have $\eta_{\scriptscriptstyle 4}=2.9280
\times 10^{-4}$ and
those to the right have $\lambda = 5.8511$\vspace{0.1in}.}
\begin{tabular}{ccc||ccc}\hline
$\eta_{\scriptscriptstyle 4}$ & $\lambda$ & $Tr\, P$ &
   $\lambda$ & $\eta_{\scriptscriptstyle 4}$ & $Tr\, P$ \\\hline
$2.9280\times 10^{-4}$ & $5.8510$ & $7.22\times 10^{-5}$ &
   $5.8511$ & $2.9279\times 10^{-4}$ & $7.03\times 10^{-4}$\\
         &  $5.8511$ & $8.82\times 10^{-6}$ &
              & $2.9280\times 10^{-4}$ & $8.82\times 10^{-6}$ \\
         & $5.8512$ & $1.84\times 10^{-5}$ &
            & $2.9281\times 10^{-4}$ & $6.14\times 10^{-4}$ \\\hline
\end{tabular} 
\end{center}
\end{table}
Of the two sweep parameters, $\eta_{\scriptscriptstyle 4}$ is seen to have 
the largest impact on gate performance. This will turn out to be true for 
the other one-qubit gates as well. Although TRP can produce a non-adiabatic 
Hadamard gate whose error probability falls below the accuracy threshold 
$P_{a}\sim 10^{-4}$, it is clear from Table~\ref{table1} that the sweep 
parameters must be controlled to $5$ significant figures to achieve this 
level of performance. See Section~\ref{sec5} for further discussion.

\subsubsection*{\underline{\mbox{$V_{P}$} Gate}}
\noindent The modified phase gate $V_{P}$ (see eq.~(\ref{Vpdef})) is the 
target gate here. The sweep parameters $\lambda = 5.9750$ and 
$\eta_{\scriptscriptstyle 4} = 3.8060\times 10^{-4}$ produce a unitary gate 
$U_{a}$ \cite{lhg} which has $Tr\, P=8.20\times 10^{-5}$. This gives a fidelity 
$\calF_{V_{P}}=0.9999\; 80$ and the bound $P_{e}\leq 8.20\times 10^{-5}$. 
Table~\ref{table2} shows how $Tr\, P$ varies with small changes in $\lambda$
and $\eta_{\scriptscriptstyle 4}$. 
\begin{table}[ht!]
\begin{center}
\tcaption{\label{table2}Variation of $Tr\, P$ for the modified phase gate 
$V_{P}$ when the TRP sweep parameters are altered slightly from their best 
performance values. The columns to the left of center have 
$\eta_{\scriptscriptstyle 4}=3.8060
\times 10^{-4}$ and those to the right have $\lambda = 5.9750$\vspace{0.1in}.}
\begin{tabular}{ccc||ccc}\hline
$\eta_{\scriptscriptstyle 4}$ & $\lambda$ & $Tr\, P$ &
   $\lambda$ & $\eta_{\scriptscriptstyle 4}$ & $Tr\, P$ \\\hline
$3.8060\times 10^{-4}$ & $5.9749$ & $1.56\times 10^{-4}$ &
   $5.9750$ & $3.8059\times 10^{-4}$ & $2.29\times 10^{-3}$\\
         &  $5.9750$ & $8.20\times 10^{-5}$ &
              & $3.8060\times 10^{-4}$ & $8.20\times 10^{-5}$ \\
         & $5.9751$ & $1.43\times 10^{-4}$ &
            & $3.8061\times 10^{-4}$ & $1.88\times 10^{-3}$ \\\hline
\end{tabular} 
\end{center}
\end{table}
Again, gate performance is most sensitive to variation of 
$\eta_{\scriptscriptstyle 4}$, and
the sweep parameters must be controlled to high precision to surpass the 
accuracy threshold (see Section~\ref{sec5}).

\subsubsection*{\underline{\mbox{$V_{\pi /8}$} Gate}}
\noindent The target gate this time is the modified $\pi /8$ gate $V_{\pi /8}$
(see eq.~(\ref{Vp8def})). For $\lambda = 6.0150$ and 
$\eta_{\scriptscriptstyle 4}=8.1464\times 10^{-4}$, TRP produces a unitary gate 
$U_{a}$ \cite{lhg} which has $Tr\, P=3.03\times 10^{-5}$, fidelity 
$\calF_{V_{\pi /8}} = 0.9999\; 92$, and $P_{e}\leq 3.03\times 10^{-5}$. 
Table~\ref{table3} shows how gate performance varies when the sweep parameters 
are altered slightly.
\begin{table}[ht!]
\begin{center}
\tcaption{\label{table3}Variation of $Tr\, P$ for the modified $\pi /8$
gate when the TRP sweep parameters are altered slightly from their best 
performance values. The columns to the left of center have 
$\eta_{\scriptscriptstyle 4}=8.1464
\times 10^{-4}$ and those to the right have $\lambda = 6.0150$\vspace{0.1in}.}
\begin{tabular}{ccc||ccc}\hline
$\eta_{\scriptscriptstyle 4}$ & $\lambda$ & $Tr\, P$ &
   $\lambda$ & $\eta_{\scriptscriptstyle 4}$ & $Tr\, P$ \\\hline
$8.1464\times 10^{-4}$ & $6.0149$ & $1.30\times 10^{-3}$ &
   $6.0150$ & $8.1463\times 10^{-4}$ & $1.77\times 10^{-3}$\\
         &  $6.0150$ & $3.03\times 10^{-5}$ &
              & $8.1464\times 10^{-4}$ & $3.03\times 10^{-5}$ \\
         & $6.0151$ & $2.18\times 10^{-3}$ &
            & $8.1465\times 10^{-4}$ & $2.77\times 10^{-3}$ \\\hline
\end{tabular} 
\end{center}
\end{table}
As with the two previous gates, performance is most sensitive to variation of
$\eta_{\scriptscriptstyle 4}$, and the sweep parameters must be controllable 
to high precision (see Section~\ref{sec5}).

\subsubsection*{\underline{NOT Gate}}
\noindent Here the target gate is the NOT gate (see eq.(\ref{Udefs2})). 
For $\lambda = 7.3205$ and $\eta_{\scriptscriptstyle 4}= 2.9277\times 
10^{-4}$, TRP produces the unitary gate $U_{a}$ \cite{lhg} for which 
$Tr\, P= 1.10\times 10^{-5}$, fidelity $\calF_{NOT}=0.9999\; 97$, and 
$P_{e}\leq 1.10 \times 10^{-5}$. Table~\ref{table4} shows how small variation 
of the sweep parameters alters $Tr\, P$.
\begin{table}[ht!]
\begin{center}
\tcaption{\label{table4}Variation of $Tr\, P$ for the NOT gate when the TRP 
sweep parameters are altered slightly from their best performance values. The 
columns to the left of center have $\eta_{\scriptscriptstyle 4}=2.9277\times 
10^{-4}$ and 
those to the right have $\lambda = 7.3205$\vspace{0.1in}.}
\begin{tabular}{ccc||ccc}\hline
$\eta_{\scriptscriptstyle 4}$ & $\lambda$ & $Tr\, P$ &
   $\lambda$ & $\eta_{\scriptscriptstyle 4}$ & $Tr\, P$ \\\hline
$2.9277\times 10^{-4}$ & $7.3204$ & $1.12\times 10^{-5}$ &
   $7.3205$ & $2.9276\times 10^{-4}$ & $1.23\times 10^{-3}$\\
         &  $7.3205$ & $1.10\times 10^{-5}$ &
              & $2.9277\times 10^{-4}$ & $1.10\times 10^{-5}$ \\
         & $7.3206$ & $1.22\times 10^{-5}$ &
            & $2.9278\times 10^{-4}$ & $1.23\times 10^{-3}$ \\\hline
\end{tabular} 
\end{center}
\end{table}
As with the other one-qubit gates, performance is most sensitive to variation
in $\eta_{\scriptscriptstyle 4}$, and the sweep parameters must be 
controllable to $5$ significant figures for the gate error probability 
$P_{e}$ to fall below the accuracy threshold $P_{a}\sim 10^{-4}$ (see 
Section~\ref{sec5}).

\subsection{\label{sec3.2}Modified Controlled-Phase Gate 
$V_{\scriptscriptstyle CP}$}
\noindent We complete the TRP implementation of the universal set \uniset\ 
by showing how the modified controlled-phase gate $V_{\scriptscriptstyle CP}$
can be produced. As shown in Appendix~A, the two-qubit Hamiltonian $H_{2}
(\tau )$ used to implement $V_{\scriptscriptstyle CP}$ depends on two sets of
parameters. The first set $(\lambda$,$\,\eta_{\scriptscriptstyle 4})$ 
consists of the now familiar TRP sweep parameters, while the second set 
$(d_{1}$,$\,d_{2}$,$\, d_{3}$,$\, d_{4}$,$\, c_{4})$ consists of 
parameters such as coupling constants and frequency-related shifts and 
differences that appear in $H_{2}(\tau )$. We stress that only the first set 
are directly related to the TRP sweeps. Unlike the situation encountered in 
Section~\ref{sec3.1}, best gate performance does \textit{not\/} require high 
precision control of the TRP sweep parameters $\lambda$ and 
$\eta_{\scriptscriptstyle 4}$. Instead, the critical parameters for gate 
performance will turn out to be $d_{1}$, $d_{4}$, and $c_{4}$. As this is the 
first time we present results for $V_{\scriptscriptstyle CP}$, we also
include the unitary gate $U_{a}$ produced by our best-case choice of 
parameters.

The target gate $V_{\scriptscriptstyle CP}$ has real and imaginary parts
(see eq.~(\ref{modcp}))
\begin{eqnarray}
Re\left( V_{\scriptscriptstyle CP}\right) & = &
   \left( \begin{array}{cccc}
             1 & 0 & 0 & 0 \\
             0 & 1 & 0 & 0 \\
             0 & 0 & -1 & 0 \\
             0 & 0 & 0 & 1 
          \end{array}
   \right) \nonumber\\
Im\left( V_{\scriptscriptstyle CP}\right) & = &
   \left( \begin{array}{cccc}
             0 & 0 & 0 & 0 \\
             0 & 0 & 0 & 0 \\
             0 & 0 & 0 & 0 \\
             0 & 0 & 0 & 0 
          \end{array}
   \right) . 
\end{eqnarray}
The parameters
\begin{displaymath}
\begin{array}{lclcr}
\lambda = 5.1 & & d_{1}=11.702 & & c_{4}=5.0003\\
\eta_{\scriptscriptstyle 4} = 2.4\times 10^{-4} & & d_{2}=-2.6 & & \\
 & & d_{3}=-0.41 & & \\
 & & d_{4}=6.6650 & &
\end{array} 
\end{displaymath}
produce the gate $U_{a}$ with 
\begin{eqnarray}
Re\left( U_{a}\right) & = &
   \left( \begin{array}{rrrr}
             0.9998 & 0.0155 & 0.0041 & 0.0028 \\
             -0.0154 & 0.9997 & -0.0003 & 0.0021 \\
             0.0042 & -0.0002 & -0.9999 & -0.0038 \\
             -0.0026 & -0.0021 & -0.0037 & 0.9999 
          \end{array}
   \right) \nonumber\\
Im\left( U_{a}\right) & = &
   \left( \begin{array}{cccc}
             0.0052 & -0.0108 & -0.0031 & -0.0017 \\
             -0.0109 & 0.0064 & -0.0084 & 0.0068 \\
             0.0030 & 0.0084 & 0.0060 & -0.0079 \\
             -0.0018 & 0.0068 & 0.0079 & 0.0026 
          \end{array}
   \right) . 
\end{eqnarray}
All two-qubit simulations used a dimensionless inversion time $\tau_{0}
= 120.00$. From $U_{a}$ and $V_{\scriptscriptstyle CP}$ it follows that
$Tr\, P = 1.27\times 10^{-3}$, the gate fidelity $\calF_{V_{CP}}=0.9996\: 83$,
and $P_{e}\leq 1.27\times 10^{-3}$. Table~\ref{table5} shows how gate
performance varies when either $\lambda$ or $\eta_{\scriptscriptstyle 4}$ is
altered slightly. 
\begin{table}[ht!]
\begin{center}
\tcaption{\label{table5}Variation of $Tr\, P$ for the modified 
controlled-phase gate $V_{\scriptscriptstyle CP}$ when the TRP sweep 
parameters are altered slightly from their best performance values. The 
columns to the left (right) of center vary $\lambda$ 
($\eta_{\scriptscriptstyle 4}$), while holding all other parameters
fixed\vspace{0.1in}.}
\begin{tabular}{cc||cc}\hline
$\lambda$ & $Tr\, P$ &
   $\eta_{\scriptscriptstyle 4}$ & $Tr\, P$ \\\hline
$5.0$ & $2.70\times 10^{-3}$ &
   $2.3\times 10^{-4}$ & $1.46\times 10^{-3}$\\
$5.1$ & $1.27\times 10^{-3}$ &
   $2.4\times 10^{-4}$ & $1.27\times 10^{-3}$ \\
$5.2$ & $2.10\times 10^{-3}$ &
   $2.5\times 10^{-4}$ & $1.35\times 10^{-3}$ \\\hline
\end{tabular} 
\end{center}
\end{table}
Notice that the TRP sweep parameters only need to be controlled to 
\textit{two\/} significant figures. A similar situation occurs for the 
parameters $d_{2}$ and $d_{3}$, though in the interests of brevity, we will 
not include Tables to show this. Table~\ref{table6} shows how gate performance 
varies when either $d_{1}$ or $d_{4}$ is varied slightly.
\begin{table}[ht!]
\begin{center}
\tcaption{\label{table6}Variation of $Tr\, P$ for the modified 
controlled-phase gate $V_{\scriptscriptstyle CP}$ when the parameters $d_{1}$
and $d_{4}$ are altered slightly from their best performance values. The 
columns to the left (right) of center vary $d_{1}$ ($d_{4}$), while holding 
all other parameters fixed\vspace{0.1in}.}
\begin{tabular}{cc||cc}\hline
$d_{1}$ & $Tr\, P$ &
   $d_{4}$ & $Tr\, P$ \\\hline
$11.699$ & $1.41\times 10^{-2}$ &
   $6.6647$ & $1.31\times 10^{-2}$\\
$11.700$ & $7.63\times 10^{-3}$ &
   $6.6648$ & $6.35\times 10^{-3}$ \\
$11.701$ & $3.36\times 10^{-3}$ &
   $6.6649$ & $2.40\times 10^{-3}$ \\
$11.702$ & $1.27\times 10^{-3}$ &
   $6.6650$ & $1.27\times 10^{-3}$ \\
$11.703$ & $1.43\times 10^{-3}$ &
   $6.6651$ & $2.97\times 10^{-3}$ \\
$11.704$ & $3.79\times 10^{-3}$ &
   $6.6652$ & $7.59\times 10^{-3}$ \\
$11.705$ & $8.27\times 10^{-3}$ &
   $6.6653$ & $1.50\times 10^{-2}$ \\\hline
\end{tabular} 
\end{center}
\end{table}
We see that these parameters need to be controlled to five significant figures,
although a small amount of uncertainty in the fifth significant figure will
not drastically damage performance. Recall from Appendix~A that $d_{1}$ is the 
dimensionless version of the difference in qubit Larmor frequencies, and 
$d_{4}$ is the dimensionless Ising coupling constant. Finally, 
Table~\ref{table7} shows how $Tr\, P$ varies when $c_{4}$ changes slightly.
\begin{table}[ht!]
\begin{center}
\tcaption{\label{table7}Variation of $Tr\, P$ for the modified 
controlled-phase gate $V_{\scriptscriptstyle CP}$ when the parameter $c_{4}$ 
is varied slightly from its best performance value. The columns to the left 
(right) of center vary $c_{4}$ in the fifth (fourth) significant figure, while 
holding all other parameters fixed\vspace{0.1in}.}
\begin{tabular}{cc||cc}\hline
$c_{4}$ & $Tr\, P$ &
   $c_{4}$ & $Tr\, P$ \\\hline
$5.0000$ & $1.98\times 10^{-3}$ &
   $4.999$ & $1.50\times 10^{-2}$\\
$5.0001$ & $1.55\times 10^{-3}$ &
   $5.000$ & $1.98\times 10^{-3}$ \\
$5.0002$ & $1.36\times 10^{-3}$ &
   $5.001$ & $5.48\times 10^{-3}$ \\
$5.0003$ & $1.27\times 10^{-3}$ &
   &  \\
$5.0004$ & $1.38\times 10^{-3}$ &
   &  \\
$5.0005$ & $1.65\times 10^{-3}$ &
   &  \\
$5.0006$ & $2.11\times 10^{-3}$ &
   &  \\\hline
\end{tabular} 
\end{center}
\end{table}
We see that best performance is not seriously compromised if $c_{4}$ has a 
small uncertainty in its fifth significant figure. Interestingly, if
$c_{4}$ can be controlled to four significant figures, we can come
very close to best performance. As shown in Appendix~A, $c_{4}$ is a 
dimensionless degeneracy-breaking parameter. Although gate performance for 
$V_{\scriptscriptstyle CP}$ is slightly more robust than for the one-qubit 
gates of Section~\ref{sec3.1}, we have not yet been able to find a combination 
of sweep parameters and two-qubit interaction that yields $P_{e}< 10^{-4}$.
See Section~\ref{sec5} for further discussion.

\section{\label{sec4}Realizing TRP in Superconducting Qubit Systems}
\noindent Here we demonstrate how TRP sweeps can be applied to superconducting 
(SC) qubit systems \cite{makh,nori}. We first consider a SC charge qubit in 
Section~\ref{sec4.1}, then go on to flux qubits in Section~\ref{sec4.2}. Two 
realizations of a flux qubit are considered: first the rf-SQUID qubit in 
Section~\ref{sec4.2.1}, then the persistent-current qubit in 
Section~\ref{sec4.2.2}. For each of these qubit realizations, the 
demonstration proceeds in $3$ steps: (i)~the appropriate one-qubit Hamiltonian 
is introduced; (ii)~the coefficients of $\sigma_{z}$ and $\sigma_{x}$ in 
this Hamiltonian are identified with $at$ and $b\,\cos\phi_{trp}(t)$, 
respectively; and (iii)~the rotating-wave approximation is invoked to obtain 
the one-qubit TRP Hamiltonian (see eqs.~(\ref{zeeman}) and (\ref{TRPsweep})). 
In each demonstration, the second step establishes the link between the 
theoretical parameters ($a, b, \phi_{trp}(t)$) and the experimental control 
fields acting on the SC qubit. 

\subsection{\label{sec4.1}Superconducting Charge Qubit}
\noindent Figure~\ref{fig1} shows a quantum circuit that can act as a charge 
qubit with adjustable Josephson coupling \cite{aver,makh2,fn2}.
\setlength{\unitlength}{0.025in}
\begin{figure}[htbp]
\begin{center}
\begin{picture}(200,90)(-32.5,0)
\put(80,60){$\times$}
\put(66,60){$\scriptstyle E_{J}^{0},C_{J}$}
\put(95,60){$\Phi_{x}$}
\put(110,60){$\times$}
\put(115.5,60){$\scriptstyle E_{J}^{0},C_{J}$}
\put(80,45){\framebox(33.5,10){{\scriptsize SC Island}}}
\put(82.0508,55){\line(0,1){15}}
\put(112.0508,55){\line(0,1){15}}
\put(82.0508,70){\line(1,0){30}}
\put(96.75,70){\line(0,1){10}}
\put(40,80){\line(1,0){56.75}}
\put(96.75,38){\line(0,1){7}}
\put(91.75,38){\line(1,0){10}}
\put(91.75,34.5){\line(1,0){10}}
\put(104,35){$\scriptstyle C_{g}$}
\put(96.75,26){\line(0,1){8.5}}
\put(89.5,26){\line(1,0){15}}
\thicklines
\put(93,23){\line(1,0){7.5}}
\thinlines
\put(106,22.5){$\scriptstyle V_{g}$}
\put(96.75,13){\line(0,1){10}}
\put(40,13){\line(1,0){56.75}}
\put(40,13){\line(0,1){67}}
\end{picture}
\end{center}
\fcaption{\label{fig1}Quantum circuit for a charge qubit with tunable Josephson
coupling. The superconducting (SC) island is connected to a SC  electrode 
through a dc-SQUID that is threaded by an external magnetic flux $\Phi_{x}$. 
Each cross $\times$ in the dc-SQUID represents a Josephson junction. The 
island is also coupled to a gate voltage $V_{g}$ through a capacitor $C_{g}$.}
\end{figure}
The circuit consists of a superconducting (SC) island that is connected to a
SC electrode via a dc-SQUID that is threaded by an external magnetic flux
$\Phi_{x}$, as well as to a gate voltage $V_{g}$ through a capacitor $C_{g}$.
The dc-SQUID is designed to have low self-inductance and is made up of two 
identical Josephson junctions (JJ), each with Josephson coupling energy 
$E_{J}^{0}$ and capacitance $C_{J}$. The SC energy gap $\Delta$ is assumed to 
be the largest energy scale for the circuit dynamics so that at sufficiently 
low temperature only Cooper pairs can tunnel through the JJs. Denoting the 
number operator for Cooper pairs on the island by $\hat{n}$ and the phase of 
the SC order parameter by $\theta$, the Hamiltonian for the SC island is
\begin{equation}
H = 4E_{c}\left(\hat{n}-n_{g}\right)^{2} - E_{J}(\Phi_{x})\cos\theta .
\label{chrgHam1}
\end{equation}
Here: (i)~$n_{g}=C_{g}V_{g}/2e$; (ii)~$E_{c}\equiv 0.5e^{2}/(C_{g}+2C_{J})$
is the charging energy; and (iii)~the flux-controlled Josephson coupling
energy is $E_{J}(\Phi_{x}) =2E_{J}^{0}\cos\left(\pi\Phi_{x}/\Phi_{0}\right)$,
where $\Phi_{0}=h/2e$ is the flux quantum. The number operator $\hat{n}$ and
the phase $\theta$ are canonically conjugate variables subject to the 
uncertainty relation $\Delta n\Delta\theta\geq 1$ \cite{and}. In the charging
limit ($E_{c}\gg E_{J}^{0}$) $H$ is dominated by the charging term in
eq.~(\ref{chrgHam1}). In this case the eigenstates $|n\rangle$ of $\hat{n}$
form a convenient basis with which to represent $H$. Because of Cooper pair 
tunneling, the Josephson coupling term in $H$ will have matrix elements 
connecting the states $|n\rangle\leftrightarrow |n+1\rangle$. It follows from
these remarks that $H$ can be written as
\begin{equation}
H = \sum_{n}\left\{\, 4E_{c}\left( n-n_{g}\right)^{2}|n\rangle\langle n|
     -\frac{E_{J}(\Phi_{x})}{2}\left[\, |n\rangle\langle n+1| +
        |n+1\rangle\langle n|\,\right]\,\right\} .
\label{chrgHam2}
\end{equation}
Note that in the absence of the Josephson coupling term, and for $n_{g}=1/2$, 
the states $|0\rangle$ and $|1\rangle$ have degenerate energies: $E_{0}=E_{1}=
E_{c}$. In fact, a level-crossing occurs at this value of $n_{g}$. The 
presence of the Josephson coupling in $H$ causes an avoided crossing to occur 
at $n_{g}=1/2$, opening an energy gap $\Delta E_{01}= E_{J}(\Phi_{x})\ll 
E_{c}$. Near the degeneracy point ($n_{g}=1/2$), these two states have the 
smallest energy eigenvalues. Thus at low temperature, and for low frequency 
gate voltage $V_{g}(t)$ and external magnetic flux $\Phi_{x}(t)$, the circuit 
dynamics near the degeneracy point is restricted to the subspace spanned by 
the states $|0\rangle$ and $|1\rangle$. Under these conditions, the only terms
in $H$ that are dynamically relevant are those that act on this subspace. 
Truncating $H$ so that only those terms are kept gives the SC charge qubit 
Hamiltonian 
\begin{equation}
H_{cq} = -\frac{1}{2}\bfsig\cdot\bfB ,
\label{chrgHam}
\end{equation}
where
\begin{eqnarray}
B_{z} & = & 4E_{c}\left( 1-2n_{g}\right) \nonumber\\
B_{x} & = & E_{J}(\Phi_{x}) \nonumber\\
B_{y} & = & 0 .
\label{Bfldcomps}
\end{eqnarray}
This completes the first step of the demonstration.

To apply a TRP sweep to a charge qubit we must require that
\begin{eqnarray}
\frac{B_{z}}{2} & = & at \nonumber\\
\frac{B_{x}}{2} & = & b\,\cos\phi_{trp} ,
\label{swpap1}
\end{eqnarray}
where $\phi_{trp}=(2/n)\, Bt^{n}$ is the TRP twist profile. This is step~2
in the demonstration. With these
assignments, eq.~(\ref{chrgHam}) becomes
\begin{equation}
H_{cq} = -at\,\sigma_{z} -b\cos\phi_{trp}\,\sigma_{x} .
\end{equation}
In the rotating wave approximation (step~3) this becomes
\begin{equation}
H_{cq} = -at\,\sigma_{z} -b\cos\phi_{trp}\,\sigma_{x} -b\sin\phi_{trp}\,
            \sigma_{y} . 
\label{cqTRPHam}
\end{equation}
Comparing this with eqs.~(\ref{zeeman}) and (\ref{TRPsweep}), we see that 
eq.~(\ref{cqTRPHam}) is the Hamiltonian for a qubit interacting with a TRP
sweep. Plugging eqs.~(\ref{Bfldcomps}) into eqs.~(\ref{swpap1}), and recalling
that $E_{J}(\Phi_{x})=2E_{J}^{0}\cos (\pi\Phi_{x}/\Phi_{0})$ gives
\begin{eqnarray}
V_{g}(t) & = & \frac{e}{C_{g}}\left(\, 1 -\frac{at}{2E_{c}}\,\right)
                \label{Vgtimedep}\\
\Phi_{x}(t) & = & \left(\frac{\Phi_{0}}{\pi}\right)\,\phi_{trp}(t)
                    \label{phixtimedep}\\
E_{J}^{0} & = & b . \label{bEjlink}
\end{eqnarray}
Eqs.~(\ref{Vgtimedep}) and (\ref{phixtimedep}) specify the time dependence
that $V_{g}(t)$ and $\Phi_{x}(t)$ must have, respectively, for a TRP sweep
to be applied to a SC charge qubit, and eq.~(\ref{bEjlink}) links the sweep
parameter $b$ to the Josephson coupling energy $E_{J}^{0}$. The TRP sweep
parameters $a$, $B$, and $T_{0}$ can then be found from eq.~(\ref{paramdefs})
for given values of $\lambda$, $\eta_{\scriptscriptstyle 4}$, and $\tau_{0}$.
Varying the gate voltage $V_{g}(t)$ according to eq.~(\ref{Vgtimedep}) 
produces the inversion of the TRP control field $\bfF (t)$ in 
eq.~(\ref{TRPsweep}), while varying the flux $\Phi_{x}(t)$ through the 
dc-SQUID according to eq.~(\ref{phixtimedep}), together with 
eq.~(\ref{bEjlink}) and the rotating-wave approximation, produces its twisting
in the $x$-$y$ plane.

\subsection{\label{sec4.2}Superconducting Flux Qubit}
\noindent We examine two flux qubit proposals: (i)~the rf-SQUID qubit 
(Section~\ref{sec4.2.1}); and (ii)~the persistent-current qubit
(Section~\ref{sec4.2.2}).

\subsubsection{\label{sec4.2.1}rf-SQUID Qubit}
\noindent In an rf-SQUID a single Josephson junction (JJ) interrupts a
superconducting (SC) loop that is threaded by an external magnetic flux
$\Phi_{x}$. Because the loop has non-vanishing self-inductance $L$, a
secondary flux $\Phi_{s}=Li_{s}$ is present whenever a supercurrent $i_{s}$
circulates around the loop. The total flux $\Phi =\Phi_{x}+\Phi_{s}$ 
determines: (i)~the phase difference $\varphi$ across the JJ via $\varphi =2\pi
\,\Phi /\Phi_{0} \pmod{2\pi}$; and (ii)~the supercurrent $i_{s}$ via $i_{s}=
i_{c}\,\sin (2\pi\Phi /\Phi_{0})$. Here $i_{c}$ is the critical current for 
the junction, and $\Phi_{0}=h/2e$ is the flux quantum. The potential energy 
for the rf-SQUID is the sum of the Josephson coupling energy and the magnetic 
energy associated with the secondary flux $\Phi_{s}=\Phi -\Phi_{x}$:
\begin{equation}
U(\Phi ) = -E_{J}^{0}\cos\left(2\pi\frac{\Phi}{\Phi_{0}}\right) +
             \frac{\left(\Phi -\Phi_{x}\right)^{2}}{2L} ;
\label{potendw}
\end{equation}
here $E_{J}^{0}$ is the coupling energy. $U(\Phi )$ has a number of 
important properties. First, $U(\Phi )$ forms a double-well potential for
$\Phi$-values near $\Phi_{0}/2$ when: (i)~$\Phi_{x}\approx\Phi_{0}/2$; and 
(ii)~the self-inductance $L$ is large enough to cause $\beta_{L}=E_{J}^{0}/
(\Phi_{0}^{2}/4\pi^{2}L)>1$. The minima occur at $\Phi_{\pm}=\Phi_{0}/2 \pm
\delta\Phi_{\pm}$, and the supercurrents corresponding to these minima, 
$i_{\pm}=\mp i_{c}\sin (2\pi\delta\Phi_{\pm} /\Phi_{0})$, have opposite 
circulations. When $\Phi_{x}=\Phi_{0}/2$, the double-well potential is 
symmetric; the central barrier has a maximum at $\Phi=\Phi_{0}/2$, and the
barrier height at maximum is $E_{J}^{0}$. When $\Phi_{x}= \Phi_{0}/2+\delta
\Phi_{x}$, the double-well potential is asymmetric, and for small $\delta
\Phi_{x}$, the barrier height at maximum remains of order $E_{J}^{0}$.

The quantum degree of freedom for an rf-SQUID is the total flux $\Phi$,
or equivalently, the supercurrent $i_{s}$. In the absence of tunneling through
the central barrier, the groundstate is doubly degenerate and the energy
eigenstates $|L\rangle$ and $|R\rangle$ are localized, respectively, about
the left and right minima of the double-well potential. Tunneling splits the
degeneracy and the new energy eigenstates $|E_{s}\rangle$ and $|E_{a}\rangle$
are, respectively, symmetric and antisymetric linear combinations of 
$|L\rangle$ and $|R\rangle$. At sufficiently low temperature the rf-SQUID is
limited to the subspace of states spanned by $|E_{s}\rangle$ and $|E_{a}
\rangle$. Bocko et al. \cite{bocko} were the first to propose using a 
low-temperature rf-SQUID as a qubit. The computational basis states (CBS) $|0
\rangle$ and $|1\rangle$ are identified with the states $|R\rangle$ and $|L
\rangle$, respectively, and are defined to be eigenstates of the operator
$\sigma_{z}$: $\sigma_{z}|i\rangle = (-1)^{i}|i\rangle$ with $i=0,1$. For 
$\Phi_{x}\approx\Phi_{0}/2$, the asymmetry of the double-well potential 
introduces a bias $\epsilon$ in the minima of the double-well:
\begin{equation}
\epsilon = U(\Phi_{-})-U(\Phi_{+}) = 4\pi E_{J}^{0}\sqrt{6(\beta_{L}-1)}\left(
             \frac{\Phi_{x}}{\Phi_{0}}-\frac{1}{2}\right) .
\label{bias}
\end{equation}
In the absence of tunneling, the energy $E_{0}$ ($E_{1})$) of the state 
$|R\rangle =|0\rangle$ ($|L\rangle =|1\rangle$) is well approximated by the 
sum of $U(\Phi_{+})$ ($U(\Phi_{-})$) and the groundstate energy of a harmonic 
oscillator with frequency determined by the double-well curvature at $\Phi =
\Phi_{+}$ ($\Phi =\Phi_{-}$). For small asymmetry, the energy difference 
$E_{1}-E_{0}$ for the CBS is then $\epsilon$. To account for this, the flux
qubit Hamiltonian $H_{f1}$ includes the term $-(\epsilon /2)\,\sigma_{z}$, 
where the zero of energy has been set at $(E_{0}+E_{1})/2$. Note that this 
term in $H_{f1}$ can be varied by varying the flux $\Phi_{x}$ which 
alters the asymmetry of the double-well, and thus the bias $\epsilon$. 
Tunneling between wells causes a bit-flip operation $|0\rangle\leftrightarrow 
|1\rangle$ to be applied to the qubit state. If $(\beta_{L}-1)\ll 1$, 
eq.~(\ref{bias}) indicates that $|\epsilon |\ll E_{J}^{0}$, and so the 
energies $\mp \epsilon /2$ of the two lowest energy eigenstates are well below 
the barrier maximum. WKB theory can thus be used to determine the amplitude 
$t_{lr}/2$ for the qubit to tunnel from the left to the right well (the factor 
of $1/2$ is introduced for convenience). On the basis of the above remarks, we 
are led to the flux qubit Hamiltonian
\begin{equation}
H_{f1} = -\frac{\epsilon}{2}\,\sigma_{z} -\frac{t_{lr}}{2}\,\sigma_{x}
        = -\frac{1}{2}\,\bfsig\cdot\bfB ,
\label{fluxHam}
\end{equation}
where
\begin{eqnarray}
B_{z} & = & \epsilon \nonumber\\
B_{x} & = & t_{lr} \nonumber\\
B_{y} & = & 0 .
\label{rfBdefs}
\end{eqnarray}
The tunneling amplitude $t_{lr}/2$ is highly sensitive to the barrier height
which, as noted earlier, is of order $E_{J}^{0}$. If the JJ in the rf-SQUID is
replaced by a dc-SQUID threaded by an external flux $\tilde{\Phi}_{x}$ (as
in Section~\ref{sec4.1}), the coupling energy becomes adjustable, $E_{J}^{0}
\rightarrow E_{J}(\tilde{\Phi}_{x})=2E_{J}^{0}\cos (\pi\tilde{\Phi}_{x}/
\Phi_{0})$, and $\tilde{\Phi}_{x}$ can be used to vary $B_{x}=t_{lr}$. With
this modification, both terms in $H_{f1}$ are experimentally controllable. 
This leads us to the quantum circuit in Figure~\ref{fig2} for an rf-SQUID flux 
qubit. The JJs in\hspace{-0.5em} 
\setlength{\unitlength}{0.025in}
\begin{figure}[htbp]
\begin{center}
\begin{picture}(120,60)(-10,0)
\put(32,30){$\Phi_{x}$}
\put(80,30){$\tilde{\Phi}_{x}$}
\put(65,38){$\times$}
\put(54,38){$\scriptstyle E_{J}^{0},C$}
\put(95,38){$\times$}
\put(100,38){$\scriptstyle E_{J}^{0},C$}
\put(67.125,25){\line(0,1){22}}
\put(97.125,25){\line(0,1){22}}
\put(67.125,25){\line(1,0){30}}
\put(67.125,47){\line(1,0){30}}
\put(82.125,15){\line(0,1){10}}
\put(82.125,47){\line(0,1){10}}
\put(20,15){\line(0,1){42}}
\put(20,15){\line(1,0){62.125}}
\put(20,57){\line(1,0){62.125}}
\end{picture}
\fcaption{\label{fig2}Quantum circuit for an rf-SQUID flux qubit with tunable 
Josephson coupling. The larger superconducting loop is threaded by an external 
magnetic flux $\Phi_{x}$, while the smaller loop in the dc-SQUID is threaded 
by the external flux $\tilde{\Phi}_{x}$. Each cross $\times$ in the dc-SQUID 
represents a Josephson junction with coupling energy $E_{J}^{0}$ and 
capacitance $C$.}
\end{center}
\end{figure}
the dc-SQUID are identical and have coupling energy $E_{J}^{0}$
and capacitance $C$. The WKB expression for $t_{lr}/2$ is \cite{tpo}
\begin{equation}
\frac{t_{lr}}{2}  =  \frac{\hbar\omega_{\ast}}{2\pi}\,\exp\left[\, 
                      -\frac{1}{\hbar}\int_{\Phi_{-}}^{\Phi_{+}}
                        df\,\sqrt{2M_{nn}|E-U|}\,\right] .
\label{tunnamp}
\end{equation}
Here $\omega_{\ast}$ is the attempt frequency; $f=2\pi\left[ (\Phi /\Phi_{0})-
(1/2)\right]$; $M_{nn}=2C$; and $E$ is the energy. Following Ref.~\cite{tpo} 
we find that
\begin{equation}
\omega_{\ast} = \sqrt{\frac{\beta_{L}-1}{LC}} ,
\label{attempt}
\end{equation}
and denoting the argument of the exponential in eq.~(\ref{tunnamp}) by $I$, we
find that
\begin{equation}
I = \frac{8\sqrt{LC}}{\hbar}\left(\beta_{L}-1\right)^{3/2}
       E_{J}\left(\tilde{\Phi}_{x}\right) .
\label{tunint}
\end{equation}
Eqs.~(\ref{fluxHam})--(\ref{tunint}) complete the first step of our 
demonstration. 

To produce a TRP sweep the time-dependence of $\capfx$ and $\tilfx$
must be such that
\begin{eqnarray}
at & = & \frac{B_{z}}{2} \label{rfat}\\
b\,\cos\phi_{trp} & = & \frac{B_{x}}{2} \label{rfbcos} .
\end{eqnarray} 
In principle, this completes step~2 of the demonstration. In actuality,
a bit more work is needed to fully establish the link between the theory 
parameters and the control fluxes appearing in Figure~\ref{fig2}. To that end,
substituting for $B_{z}$ using eqs.~(\ref{bias}) and (\ref{rfBdefs}) in
eq.~(\ref{rfat}) gives
\begin{equation}
at = 2\pi\left\{\,\frac{\capfx}{\capfz}-\frac{1}{2}\,\right\}\,\left[\,
      E_{J}(\tilfx )\sqrt{6(\beta_{L}-1)}\,\right] ,
\label{ateq1}
\end{equation}
where $E_{J}^{0}\rightarrow E_{J}(\tilfx )$ due to the dc-SQUID in 
Figure~\ref{fig2}, and $\beta_{L}=E_{J}(\tilfx )/\left(\capfz^{2}/4\pi^{2}L
\right)$. It will be seen below that eq.~(\ref{rfbcos}) requires $\tilfx$ to
be time-dependent, and yet we would like the term in the square bracket in
eq.~(\ref{ateq1}) to be constant. To achieve this we write
\begin{equation}
\tilfx = \tilfxz +\deltilfx ,
\label{tildefxdef}
\end{equation}
where $\tilfxz$ is time-independent and will be specified below, and we
require that $\pi\deltilfx /\capfz\ll 1$. Then we can write
\begin{equation}
E_{J}(\tilfx )  =  E_{J}(\tilfxz ) +\delta E_{J} \label{EJvardef}\\
\end{equation}
and
\begin{equation}
\beta_{L}  =  \beta_{L}^{0}+\delta\beta_{L} \label{bLvardef},
\end{equation} 
where
\begin{eqnarray}
E_{J}(\tilfxz ) = 2E_{J}^{0}\cos\left(\frac{\pi\tilfxz}{\capfz}\right)
    \hspace{0.15in} & ; & \hspace{0.15in}
 \delta E_{J} = -E_{J}(\tilfxz )\tan\left(\frac{\pi\tilfxz}{\capfz}\right)
                 \,\left(\frac{\pi\deltilfx}{\capfz}\right)
       \label{EJbitdefs}
\end{eqnarray}
and
\begin{eqnarray}
{}\hspace{-0.625in}
\beta_{L}^{0} =
  \frac{E_{J}(\tilfxz )}{\left(\frac{\capfz^{2}}{4\pi^{2}L}\right)} 
          \hspace{0.15in} & ; & \hspace{0.15in}
 \delta\beta_{L} = 
   \frac{\delta E_{J}}{\left(\frac{\capfz^{2}}{4\pi^{2}L}\right)} 
       \label{bLbitdefs} .
\end{eqnarray}
We now require that $|\delta E_{J}|\ll E_{J}(\tilfxz )$ and $\delta\beta_{L}
\ll \beta_{L}^{0}-1$ so that eq.~(\ref{ateq1}) can be written as
\begin{equation}
at = 2\pi\left\{\,\frac{\capfx}{\capfz}-\frac{1}{2}\,\right\}\,
       \left[\, E_{J}(\tilfxz )\sqrt{6(\beta_{L}^{0}-1)}\,\right] .
\label{ateq2}
\end{equation}
Solving for $\capfx (t)$ gives
\begin{equation}
\capfx (t) = \frac{\capfz}{2}\left[\, 1 + 
            \frac{at}{\pi E_{J}(\tilfxz )\sqrt{6(\beta_{L}^{0}-1)}}\,\right] .
\label{phitimedep}
\end{equation}
By requiring that the flux $\capfx (t)$ satisfy eq.~(\ref{phitimedep}), 
we insure that the $\sigma_{z}$ term in $H_{f1}$ has the appropriate $at$
coefficient needed for a TRP sweep. Thus appropriate variation of the flux 
$\capfx (t)$ through the primary loop in Figure~\ref{fig2} produces the 
inversion of the z-component of the control field $\bfF (t)$ in 
eq.~(\ref{TRPsweep}). Next we use eq.~(\ref{rfbcos}) to determine the flux 
$\tilfx (t)$ through the dc-SQUID in Figure~\ref{fig2}. Using 
eqs.~(\ref{rfBdefs})--(\ref{tunint}) in eq.~(\ref{rfbcos}) gives 
\begin{equation}
b\,\cos\phi_{trp} = \frac{\hbar}{2\pi}\,\omega_{\ast}\,e^{-I} .
\label{bcoseq1}
\end{equation}
Writing $I=I_{0}+\delta I$, where $\delta I\ll I$ and
\begin{eqnarray}
I_{0} & = & \frac{8\sqrt{LC}}{\hbar}\left(\beta_{L}^{0}-1\right)^{3/2}
             E_{J}(\tilfxz ) \label{Izdef} \\
\delta I & = & I_{0}\,\left[\frac{3}{2}
               \frac{\delta\beta_{L}}{(\beta_{L}^{0}-1)} +
           \frac{\delta E_{J}}{E_{J}(\tilfxz )}\right] , \label{delIzdef}
\end{eqnarray}
allows eq.~(\ref{bcoseq1}) to be written as
\begin{equation}
b\,\cos\phi_{trp} = \frac{\hbar}{2\pi}\,\omega_{\ast}^{0}\, e^{-I_{0}}\,
       \left[\, 1 - I_{0}\left(\,\frac{3}{2}
        \frac{\delta\beta_{L}}{(\beta_{L}^{0}-1)} +
         \frac{\delta E_{J}}{E_{J}(\tilfxz )}\,\right)\,\right] .
\label{bcoseq2}
\end{equation}
Note that $\omega_{\ast}=\omega_{\ast}^{0}+\delta \omega_{\ast}$ has been 
written as $\omega_{\ast}^{0}$ in eq.~(\ref{bcoseq2}), where
\begin{equation}
\omega_{\ast}^{0} = \sqrt{\frac{\beta_{L}^{0}-1}{LC}} .
\label{omegzdef}
\end{equation}
The reason for this is that the $\delta I$ contribution to eq.~(\ref{bcoseq2}) 
dominates the $\delta\omega_{\ast}$ contribution, and so the latter 
contribution can be safely discarded. Using eqs.~(\ref{EJbitdefs}) and 
(\ref{bLbitdefs}) in eq.~(\ref{bcoseq2}), along with some algebra, gives
\begin{equation}
\left(\, b\,\cos\phi_{trp}\,\right)\left[\,\frac{2\pi}{\hbar\omega_{\ast}^{0}}
 \, e^{I_{0}}\right] = 1 - \frac{I_{0}}{2}
  \frac{(5\beta_{L}^{0}-2)}{(\beta_{L}^{0}-1)}
   \frac{\delta E_{J}}{E_{J}(\tilfxz )} .
\label{bcoseq3}
\end{equation}
Finally, using eq.~(\ref{EJbitdefs}) again, and solving for 
$\pi\deltilfx /\capfz$ gives
\begin{equation}
\frac{\pi\deltilfx}{\capfz} = C\,\cos\phi_{trp} - D ,
\label{delphixres}
\end{equation}
where
\begin{eqnarray}
C & = & \frac{1}{8\sqrt{LC}}\left(\frac{b}{\hbar}\right)
         \left(\frac{\hbar}{E_{J}^{0}}\right)
          \left(\frac{2\pi}{\omega_{\ast}^{0}}\right)
\left[
\frac{\csc\frac{\pi\tilfxz}{\capfz}}{(5\beta_{L}^{0}-2)\sqrt{\beta_{L}^{0}-1}}
 \right]\, \exp\left[\, I_{0}\,\right] \nonumber\\
D & = & \left(\frac{1}{8\sqrt{LC}}\right)\left( 
\frac{\csc\frac{\pi\tilfxz}{\capfz}}{(5\beta_{L}^{0}-2)\sqrt{\beta_{L}^{0}-1}}
 \right)\left(\frac{\hbar}{E_{J}^{0}}\right) .
\label{CDdefs}
\end{eqnarray}
Typical values for the parameters in eqs.~(\ref{CDdefs}) are: $\beta_{L}^{0}-1=
0.1$; $\sqrt{LC}=1\,$ns; and $E_{J}^{0}/\hbar = 100\,\mathrm{GHz}$.
With these values, and choosing $b/\hbar =400\mathrm{Hz}$, and 
\begin{equation}
\tilfxz /\capfz = 1/2-\epsilon /\pi ,
\label{assignval}
\end{equation} 
with $\epsilon =0.25$ gives
\begin{eqnarray*}
\omega_{\ast}^{0} & = & 0.32\,\mathrm{GHz}\\
I_{0} & = & 12.7 .
\end{eqnarray*}
From these values we find that
\begin{eqnarray*}
C & = & 2.9\times 10^{-3}\\
D & = & 1.1\times 10^{-3} ,
\end{eqnarray*}
and from eq.~(\ref{delphixres}), it follows that $\pi\deltilfx /\capfz\approx 
10^{-3}\ll 1$ as required. The TRP sweep parameters $a$, $B$, and $T_{0}$ are 
then determined from eq.~(\ref{paramdefs}) for given values of $\lambda$,  
$\eta_{\scriptscriptstyle 4}$, and $\tau_{0}$. Thus, with $\capfx$ given by 
eq.~(\ref{phitimedep}) and $\tilfx$ given by eqs.~(\ref{tildefxdef}), 
(\ref{delphixres}), and (\ref{assignval}), we arrive at the flux qubit 
Hamiltonian is
\begin{displaymath}
H_{f1} = -at\,\sigma_{z}-b\,\cos\phi_{trp}\,\sigma_{x} .
\end{displaymath}
We see that $\capfx (t)$ produces the inversion in the TRP control field
$\bfF (t)$ in eq.~(\ref{TRPsweep}), while $\tilfx (t)$ will be seen 
momentarily to give rise to its twisting in the $x$-$y$ plane. This finally 
completes step~2. In the rotating wave approximation (step~3), $H_{f1}$ 
becomes
\begin{equation}
H_{f1} = -at\,\sigma_{z}-b\,\cos\phi_{trp}\,\sigma_{x}-b\,\sin\phi_{trp}\,
           \sigma_{y} ,
\end{equation}
which is the Hamiltonian for a qubit acted on by a TRP sweep 
(eqs.~(\ref{zeeman}) and (\ref{TRPsweep})).

\subsubsection{\label{sec4.2.2}Persistent-Current Qubit}
\noindent The focus here is the $4$-junction persistent-current qubit
introduced in Refs.~\cite{tpo,jemo,fn3}. The quantum circuit for this qubit 
is shown in Figure~\ref{fig3}. It consists\hspace{-0.5em}
\setlength{\unitlength}{0.025in}
\begin{figure}[htbp!]
\begin{center}
\begin{picture}(100,80)(0,-7)
\put(18,70){$V_{A}$}
\put(78,70){$V_{B}$}
\put(20,60){\line(0,1){7}}
\put(80,60){\line(0,1){7}}
\put(15,60){\line(1,0){10}}
\put(15,55){\line(1,0){10}}
\put(75,60){\line(1,0){10}}
\put(75,55){\line(1,0){10}}
\put(-10,55){$C_{gA},V_{gA}$}
\put(90,55){$C_{gB},V_{gB}$}
\put(20,40){\line(0,1){15}}
\put(80,40){\line(0,1){15}}
\put(15,30){\framebox(10,10){SC1}}
\put(75,30){\framebox(10,10){SC2}}
\put(25,35){\line(1,0){50}}
\put(50,33.75){$\times$}
\put(43,28){$E_{J3},C_{3}$}
\put(50,44){$\Phi_{2}$}
\put(40,55){\line(1,0){24}}
\put(50,53.75){$\times$}
\put(43,59){$E_{J4},C_{4}$}
\put(40,35){\line(0,1){20}}
\put(64,35){\line(0,1){20}}
\put(17.95,18){$\times$}
\put(-2,18){$E_{J1},C_{1}$}
\put(20,0){\line(0,1){30}}
\put(77.95,18){$\times$}
\put(84,18){$E_{J2},C_{2}$}
\put(80,0){\line(0,1){30}}
\put(20,0){\line(1,0){60}}
\put(50,10){$\Phi_{1}$}
\end{picture}
\fcaption{\label{fig3}Quantum circuit for a $4$-Josephson junction
persistent-current qubit. Junctions~$1$ and $2$ are identical and have coupling
energy $E_{J1}=E_{J2}=E_{J}^{0}$ and capacitance $C_{1}=C_{2}=C$. Junctions~$3$
and $4$ are also identical, with coupling energy $E_{J3}=E_{J4}=\beta 
E_{J}^{0}$ and capacitance $C_{3}=C_{4}=\beta C$. Superconducting islands SC1 
and SC2 are, respectively, connected to gate voltages $V_{A}$ and $V_{B}$ 
through gate capacitors $C_{gA}$ and $C_{gB}$, where $C_{gA}=C_{gB}=\gamma C$. 
The potential difference across each gate capacitor is $V_{gI}$, where $I=A,B$.
Finally, the dc-SQUID loop is threaded by a flux $\Phi_{2}$, while the lower 
loop is threaded by a flux $\Phi_{1}$.}
\end{center}
\end{figure}
of a loop with negligible self-inductance that is threaded by an external 
magnetic flux $\Phi_{1}$. Because of the small self-inductance, a supercurrent 
$i_{s}$ circulating around the loop produces negligible secondary flux and so 
the total flux through the loop $\Phi$ is given by the external flux 
$\Phi_{1}$. The loop is interrupted by two identical Josephson junctions~$1$ 
and $2$; a dc-SQUID containing two identical junctions~$3$ and $4$ whose loop 
is threaded by a flux $\Phi_{2}$; and two superconducting islands SC1 and SC2. 
Junctions~$1$ and $2$ both have coupling energy $E_{J}^{0}$ and capacitance 
$C$, while junctions~$3$ and $4$ have coupling energy $\beta E_{J}^{0}$ and 
capacitance $\beta C$. Superconducting island SC1 (SC2) is connected to a gate 
voltage $V_{A}$ ($V_{B}$) through a gate capacitor $C_{gA}$ ($C_{gB}$). The
gate capacitors are identical with $C_{gA}=C_{gB}=\gamma C$, and capacitor
$C_{gI}$ has a potential difference $V_{gI}$ across it, where $I=A,B$. The 
circuit is assumed to be at sufficiently low temperature that only 
supercurrents flow in it.

It is conventional to introduce the magnetic frustration $f_{i}=\Phi_{i}/
\Phi_{0}$, where $\Phi_{0}=h/2e$ is the flux quantum and $i=1,2$. Denoting the
Josephson phase difference across the $j^{\mathrm{th}}$ junction by 
$\varphi_{j}$, fluxoid quantization around the dc-SQUID loop in 
Figure~\ref{fig3} requires $\varphi_{4}-\varphi_{3}=-2\pi f_{2}$, and
around the lower loop requires $\varphi_{1}-\varphi_{2}+\varphi_{3}=
-2\pi f_{1}$. Adding up the Josephson coupling energy for each junction and
using the fluxoid quantization relations gives the total Josephson energy $U$:
\begin{equation}
\frac{U}{E_{J}^{0}} = 2+2\beta -2\cos\varphi_{p}\cos\varphi_{m}
      -\left\{ 2\beta\cos\left( \pi f_{a}\right)\right\}\cos\left( 2\pi f_{b}
        +2\varphi_{m}\right) ,
\end{equation}
where $\varphi_{p}=\left(\varphi_{1}+\varphi_{2}\right)/2$; $\varphi_{m}=
\left(\varphi_{1}-\varphi_{2}\right)/2$; $f_{a}=f_{2}$; and $f_{b}=f_{1}+
\left( f_{2}/2\right)$. The Josephson energy $U$ acts as the potential
energy for the persistent-current qubit. It is periodic in $f_{b}$ with period
$1$, and is symmetric about $f_{b}=1/2$. At the degeneracy point $f_{b}=1/2$, 
$\alpha =2\beta\cos\left(\pi f_{a}\right) = 1/2$, $U$ has degenerate minima at 
$(\varphi_{p},\varphi_{m})=(0,\pm\varphi_{m}^{\ast})$, where 
$\cos\varphi_{m}^{\ast}=1/(2\alpha )$. Because $U$ is periodic in $\varphi_{p}$
and $\varphi_{m}$, the degenerate minima trace out a lattice whose primitive
unit cell contains only one pair of degenerate minima. When $f_{b}$ is near 
$1/2$, the pair of minima inside a unit cell are no longer degenerate. Within 
a unit cell, the potential $U$ takes the form of a double well potential which 
is symmetric when $f_{b}=1/2$, and is asymmetric when $f_{b}$ is near $1/2$. 
Variation of $\alpha$ allows the potential energy landscape to be further 
modified. In particular, $\alpha$ can be used to suppress tunneling between 
minima located in different unit cells.

In the quantum limit: (i)~the fluxes $\varphi_{p}$ and $\varphi_{m}$ become 
quantum degrees of freedom with conjugate momenta $P_{p}=-i\hbar\partial /
\partial\varphi_{p}$ and $P_{m}=-i\hbar\partial /\partial\varphi_{m}$; and
(ii)~the Hamiltonian driving the quantum dynamics is
\begin{displaymath}
H = \frac{P_{p}^{2}}{2M_{p}} + \frac{P_{m}^{2}}{2M_{m}}+E_{J}^{0}\, U .
\end{displaymath}
Here $M_{p}=\left(\Phi_{0}/2\pi\right)^{2}\left( 1+\gamma\right) (2C)$ and
$M_{m}=\left(\Phi_{0}/2\pi\right)^{2}\left( 1+4\beta +\gamma\right) (2C)$.
The energy levels form bands due to the periodicity of $U$, and the bands are
symmetric about $f_{b}=1/2$. At sufficiently low temperature, and for $f_{b}$
near $1/2$, the circuit is effectively restricted to the subspace spanned by
the two lowest energy eigenstates $|E_{0}\rangle$ and $|E_{1}\rangle$.
Persistent supercurrents $i_{0}$ and $i_{1}$ flow in $|E_{0}\rangle$ and
$|E_{1}\rangle$, respectively, and $i_{0}=-i_{1}$.
These two eigenstates constitute the CBS $|0\rangle$ and $|1\rangle$, and the 
bit values are encoded into the direction of circulation of the 
persistent-currents. In the tight-binding approximation the states 
$|E_{0}\rangle$ and $|E_{1}\rangle$ are localized near the minima of $U$, and 
(as noted earlier) within a unit cell, $U$ has the form of a double well 
potential. Focusing on the unit cell containing the minima $(0,\pm
\varphi_{m}^{\ast})$, and denoting the state localized about the minimum 
$(0,+\varphi_{m}^{\ast})$ ($(0,-\varphi_{m}^{\ast})$) by $|+\rangle$ 
($|-\rangle$), inside this unit cell we have $|E_{0}\rangle =\left[ |+\rangle 
+|-\rangle\right]/\sqrt{2}$ and $|E_{1}\rangle =\left[ |+\rangle -|-\rangle
\right]/\sqrt{2}$. When $\alpha =2 \beta\cos (\pi f_{a})$ is chosen 
appropriately, tunneling can be restricted to the two minima within a unit 
cell. In the absence of tunneling, $|+\rangle$ ($|-\rangle$) is the 
groundstate of a qubit localized in the well centered about $(0,+
\varphi_{m}^{\ast})$ ($(0,-\varphi_{m}^{\ast})$). Altering $f_{b}$ alters the 
energies $E_{\pm}$ of the states $|\pm\rangle$, and so alters the bias $F=(
E_{+}-E_{-})/2$. Altering $f_{a}$ alters the barrier height, and so alters the 
tunneling amplitude $t$. Thus, in the representation spanned by $|\pm\rangle$, 
the persistent-current qubit Hamiltonian $H_{f2}$ is
\begin{displaymath}
H_{f2} = \left( \begin{array}{cc}
                   -F & -t\\
                   -t & F
                \end{array}
         \right) .
\end{displaymath}
Since the CBS are the eigenstates of $H_{f2}$, we can also write
\begin{displaymath}
H_{f2} = -\sqrt{F^{2}+t^{2}}\,\sigma_{z} ,
\end{displaymath}
where $\sigma_{z}=|0\rangle\langle 0|-|1\rangle\langle 1|$. Following
Ref.~\cite{tpo}, the quantum circuit is to operate near the degeneracy point 
at which $f_{1}^{\ast}=f_{2}^{\ast}=1/3$. The operating point is chosen to 
have $f_{1}^{o}=f_{1}^{\ast}+\epsilon_{1}$ and $f_{2}^{o}=f_{2}^{\ast}+
\epsilon_{2}$. It can be shown that, at the operating point, the bias is
$F_{o}=r_{1}\epsilon_{1}+r_{2}\epsilon_{2}$ and the tunneling amplitude is
$t_{o}=t_{\ast}+s_{2}\epsilon_{2}$, where $t_{\ast}$ is the tunneling
amplitude at the degeneracy point. Explicit formulas for $r_{1}$, $r_{2}$, and
$s_{2}$ appear in Ref.~\cite{tpo}, though they will not be needed here. 
By varying $f_{1}=f_{1}^{o}+\delta_{1}$ and $f_{2}=f_{2}^{o}+\delta_{2}$
about the operating point with time-dependent $\delta_{1}$ and $\delta_{2}$, 
a unitary transformation can be applied to the qubit. Defining $\tan\theta_{o}
=-t_{o}/F_{o}$, the Hamiltonian $H_{f2}$ becomes
\begin{equation}
H_{f2} = \left[\, -Z_{0}+Z_{1}\,\delta_{1}+Z_{2}\,\delta_{2}\,\right]\,
             \sigma_{z} \, -\,\left[\, X_{1}\delta_{1}+X_{2}\,\delta_{2}\,
              \right]\,\sigma_{x} ,
\label{Hf2_1}
\end{equation}
where
\begin{equation}
\begin{array}{lcl}
Z_{0}=\sqrt{F_{o}^{2}+t_{o}^{2}} & & \\
Z_{1}=r_{1}\cos\theta_{o} & ; & X_{1}=r_{1}\sin\theta_{o}\\
Z_{2}=\frac{r_{1}}{2}\cos\theta_{o}-s_{2}\sin\theta_{o} & ; & 
 X_{2} = \frac{r_{1}}{2}\sin\theta_{o}+s_{2}\cos\theta_{o}
\end{array} .
\label{Hf2_2}
\end{equation}
Inserting typical parameter values, $H_{f2}$ becomes \cite{tpo}:
\begin{equation}
\frac{H_{f2}}{E_{J}^{0}} = \left[\, -0.025 +  4.0\,\delta_{1}
       +2.1\,\delta_{2}\right]\,\sigma_{z} -\left[\, 0.46\,\delta_{1}+
         0.41\,\delta_{2}\,\right]\,\sigma_{x} .
\label{Hf2_3}
\end{equation}
This completes step~1 of the demonstration.

To produce a TRP sweep, we must require (step~2) that 
\begin{eqnarray}
Z_{0}-Z_{1}\,\delta_{1}-Z_{2}\,\delta_{2} & = & at \nonumber\\
{}\hspace{0.15in} X_{1}\,\delta_{1}+X_{2}\,\delta_{2} & = & b\,\cos\phi_{trp} .
\label{pcq1}
\end{eqnarray}
Recalling that $t=(b/a)\tau$, and defining $Z_{i}=E_{J}^{0}\, z_{i}$ and 
$X_{i}=E_{J}^{0}\, x_{i}$, eq.~(\ref{pcq1}) can be rewritten as 
\begin{eqnarray*}
z_{1}\,\delta_{1}+z_{2}\,\delta_{2} & = & z_{0}-\left(\frac{b}{E_{J}^{0}}
                                      \right)\,\tau \\
x_{1}\,\delta_{1}+x_{2}\,\delta_{2} & = & \left(\frac{b}{E_{J}^{0}}\right) .
\end{eqnarray*}
Solving for $\delta_{1}$ and $\delta_{2}$ gives
\begin{eqnarray}
\delta_{1} & = & \frac{x_{2}}{G}\left[\,\left(\frac{b}{E_{J}^{0}}\right)\tau
                  -z_{0}\,\right] + \frac{z_{2}}{G}\left(\frac{b}{E_{J}^{0}}
                   \right)\,\cos\phi_{trp} \nonumber\\
\delta_{2} & = & \frac{x_{1}}{G}\left[\, z_{0}-\left(\frac{b}{E_{J}^{0}}\right)
                         \tau\,\right] - \frac{z_{1}}{G}\left(
                      \frac{b}{E_{J}^{0}}\right)\,\cos\phi_{trp} ,
\label{pcq2}
\end{eqnarray}
where $G=x_{1}z_{2}-x_{2}z_{1}$. Typical values for $x_{i}$ and $z_{i}$ for
$i=1,2$ can be read off from eq.~(\ref{Hf2_3}). Inserting these values into 
eq.~(\ref{pcq2}) gives
\begin{eqnarray}
\delta_{1}(\tau ) & = & 0.61\,\left[\, z_{0}-\left(\frac{b}{E_{J}^{0}}\right)
\tau\,
                  \right] -3.1\left(\frac{b}{E_{J}^{0}}\right)\cos\phi_{trp}
                    \nonumber\\
\delta_{2}(\tau ) & = & 0.68\,\left[\, \left(\frac{b}{E_{J}^{0}}\right)\tau -
 z_{0}\,
                  \right] +5.9\left(\frac{b}{E_{J}^{0}}\right)\cos\phi_{trp}
                  .  
\label{pcq3}
\end{eqnarray}
Recall that a TRP sweep runs over times $-\tau_{0}/2\leq \tau\leq \tau_{0}/2$,
and that our simulations used $\tau_{0}/2=40, 60$. In Ref.~\cite{tpo}, 
$\delta_{1}\sim 10^{-3}$ and $\delta_{2}\sim 10^{-4}$. To produce $\delta_{1}$
and $\delta_{2}$ of this size using eq.~(\ref{pcq3}), we require that 
$b/E_{J}^{0}\sim 10^{-4}$ and $z_{0}\sim 10^{-4}$. With $E_{J}^{0}/\hbar =
100\, \mathrm{GHz}$, this requires $b/\hbar ,\; Z_{0}/\hbar\sim 10\, 
\mathrm{MHz}$. The TRP sweep parameters $a$, $B$, and $T_{0}$ are then found 
from eq.~(\ref{paramdefs}) for given values of $\lambda$, 
$\eta_{\scriptscriptstyle 4}$, and $\tau_{0}$. Thus, with $\delta_{1}$ and 
$\delta_{2}$ given by eqs.~(\ref{pcq3}), the persistent-current qubit 
Hamiltonian $H_{f2}$ becomes
\begin{displaymath}
H_{f2} = -at\,\sigma_{z} -b\,\cos\phi_{trp}\,\sigma_{x}.
\end{displaymath}
This completes step~2. In the rotating wave approximation (step~3), $H_{f2}$ 
becomes 
\begin{equation}
H_{f2}=-at\,\sigma_{z}-b\,\cos\phi_{trp}\,\sigma_{x}-b\,\sin\phi_{trp}\,
                  \sigma_{y}
\end{equation}
which is the Hamiltonian for a qubit subjected to a TRP sweep 
(eqs.~(\ref{zeeman}) and (\ref{TRPsweep})). We see that requiring the
control fluxes $\delta_{1}(\tau )$ and $\delta_{2}(\tau )$ to satisfy
eqs.~(\ref{pcq3}) causes a TRP sweep to be applied to the persistent-current
qubit. Unlike with the two previous SC qubit realizations, we see from
eq.~(\ref{pcq1}) that the inversion and twisting of the TRP control field
$\bfF (t)$ is determined by linear combinations of the control fluxes 
$\delta_{1}$ and $\delta_{2}$. In the case of the persistent-current qubit,
we cannot (in general) attribute inversion to one control field and twisting 
to the other.
 
\section{\label{sec5}Discussion}
\noindent In this paper we have presented simulation results which suggest
that TRP sweeps should be capable of implementing a universal set of quantum
gates \uniset\ that operate non-adiabatically and with high-fidelity. The
one-qubit gates in \uniset\ were seen to operate with gate error probabilities
satisfying $P_{e}<10^{-4}$, and the two-qubit modified controlled-phase gate
with $P_{e}<1.27\times 10^{-3}$. Using the rough-and-ready estimate for the 
accuracy threshold, $P_{a}\sim 10^{-4}$, we see that: (i)~the TRP-generated 
one-qubit gates have error probabilities that fall below this threshold value; 
and (ii)~the two-qubit gate comes within an order-of-magnitude of it. As was 
noted in Section~\ref{sec4}, this high level of gate performance requires that 
the TRP sweep parameters be controllable to high precision. Finding a way to 
make TRP gates more robust is a major challenge that must be overcome if these 
sweeps are to become a viable means for universal control of a quantum 
computer. To that end, we are currently exploring the consequences of 
interlacing TRP sweeps with dynamical decoupling pulses 
\cite{violly,zanard,llk} to produce an effective dynamics that preserves a 
group of symmetries $G$ of the target gate $U_{t}$. By removing the part of 
the TRP dynamics that does not commute with $G$, it is hoped that the 
resulting effective dynamics will yield gate performance that varies more 
slowly with the sweep parameters, and makes it easier for the minimization 
algorithms to carry out their work.

TRP has been experimentally realized in NMR systems \cite{zw1,trp2}, and 
these systems have allowed the controllable quantum interference effects that 
were predicted to arise from multiple passes through resonance per TRP sweep 
\cite{trp1} to be observed \cite{trp2}. A demonstration of how TRP sweeps can 
be applied to atomic systems using electric fields has also been given 
\cite{lhg}. In this paper we have shown how TRP can be applied to both 
superconducting charge and flux qubits. In such superconducting qubit systems, 
a qubit is realized by a mesoscopic circuit and the control fields are gate 
voltages and highly localized magnetic fields that are applied directly to the 
circuit. These systems thus allow a TRP sweep to be applied to an individual 
qubit, unlike in NMR systems where the sweeps are applied to an ensemble of 
qubits. In the NMR setting, variation of the rf-magnetic field amplitude over 
the ensemble causes the largest problem for high precision control of the TRP 
sweep parameters \cite{lhg}. Note that this difficulty does not arise with 
superconducting qubits. It would be interesting to work out an adaptation of 
the NMR experiment that observed the TRP quantum interference effects 
\cite{trp2} to a superconducting qubit system. Since these interference 
effects are a direct consequence of the temporal phase coherence of the qubit 
wave function, the adapted experiment would provide a new, independent 
demonstration of quantum coherence in superconducting qubit systems. Such a 
translation of the NMR experiment is currently underway. Ref.~\cite{lhg} 
described a state tomography experiment that would test the TRP gate 
simulation results by 
measuring the output density matrix $\rho_{exp}=U_{a}|\psi_{0}\rangle\langle
\psi_{0}|U_{a}^{\dagger}$ resulting from an initial state $|\psi_{0}\rangle$, 
for each of the TRP generated gates $U_{a}$ presented in Section~\ref{sec3}. 
Associated with each sweep is a target gate $U_{t}$ and a corresponding target 
density matrix $\rho_{t}=U_{t}|\psi_{0}\rangle\langle\psi_{0}|U_{t}^{\dagger}$.
Having measured $\rho_{exp}$, the gate fidelity could be calculated and 
compared with the fidelity obtained from the TRP gate simulations. There are 
now three possible physical systems where such an experiment could be carried 
out. For alternative applications of temporal phase coherence and rapid 
passage to quantum computing, see Refs.~\cite{fn4} and \cite{fn5,dyk}, 
respectively.

\nonumsection{Acknowledgements}
\noindent
M. Hoover was supported by the Illinois Louis Stokes Alliance for Minority
Participation Bridge to the Doctorate Fellowship, and F. Gaitan thanks 
T. Howell III for continued support.

\nonumsection{References}

\appendix

\noindent Here we derive the dimensionless Hamiltonian $H_{2}(\tau )$ 
introduced in Section~\ref{sec2.2}. It proves convenient to adopt the language
of NMR, although a more general discussion is possible.

Consider a two-qubit system in which each qubit is Zeeman-coupled to an
external magnetic field $\bfB (t)$, and the two qubits interact through the
Ising interaction. As noted in Section~\ref{sec2.2}, the Ising interaction
was chosen because of its simplicity, and because it is present in many
physical systems. It is a simple matter to alter the following arguments to 
include a different two-qubit interaction. Our starting point is thus the 
Hamiltonian
\begin{equation}
\frac{\overline{H}_{2}(t)}{\hbar} = -\frac{1}{2}\sum_{i=1}^{2}\,\gamma_{i}\,
             \bfsig_{i}\cdot\bfB (t) - \frac{\pi}{2}\, J\, \sigma_{1z}
              \sigma_{2z} ,
\label{Ham1}
\end{equation} 
where $\gamma_{i}$ is the gyromagnetic ratio for qubit~$i$, and $J$ is the
Ising interaction coupling constant. In the lab frame, $\bfB (t)$ has a
static component $B_{0}\,\hatbfz$ and a time-varying component $2B_{rf}\cos
\phi_{rf}(t)\,\hatbfx$. In the rotating wave approximation $\bfB (t)$ reduces
to
\begin{equation}
\bfB (t) = B_{0}\,\hatbfz + B_{rf}\cos\phi_{rf}(t)\,\hatbfx -B_{rf}\sin
            \phi_{rf}(t)\,\hatbfy .
\label{Brwa}
\end{equation}
Introducing $\omega_{i}=\gamma_{i}B_{0}$ and $\omega_{i}^{rf}=\gamma_{i}
B_{rf}$ ($i=1,2$), and inserting eq.~(\ref{Brwa}) into eq.~(\ref{Ham1}) gives
\begin{eqnarray}
\frac{\overline{H}_{2}(t)}{\hbar} & = & \sum_{i=1}^{2}\,\left[
         -\frac{\omega_{i}}{2}\sigma_{iz} -  \frac{\omega_{i}^{rf}}{2}
          \left\{\,\cos\phi_{rf}\sigma_{ix}-\sin\phi_{rf}\sigma_{iy}\right\}
               \right]  - \frac{\pi}{2}\, J\, \sigma_{1z} \sigma_{2z} .
\end{eqnarray}
Transformation to the detector frame is done via the unitary operator
$$U(t)=\exp\left[\, (i\phi_{det}(t)/2)\left(\sigma_{1z}+\sigma_{2z}\right)\,
\right] .$$ The Hamiltonian in the detector frame is then \cite{abr}
\begin{eqnarray}
\hspace{-0.35in}\frac{\mbox{\~{H}}_{2}(t)}{\hbar} & = & U^{\dagger}
   \left( \frac{\mbox{\~{H}}_{2}(t)}{\hbar}\right) U
                          -iU^{\dagger}\frac{dU}{dt} \nonumber\\
 & = & \sum_{i=1}^{2}\left[\left( -\frac{\omega_{i}}{2}+\dot{\phi}_{det}
          \right)\sigma_{iz}  -\frac{\omega_{i}^{rf}}{2}\left\{\cos\left(
           \phi_{det}-\phi_{rf}\right)\sigma_{ix}+\sin\left(\phi_{det}-
            \phi_{rf}\right)\sigma_{iy}\right\}\right]\nonumber\\ 
 & & \hspace{0.75in} -\frac{\pi}{2}\, J\, \sigma_{1z}\sigma_{2z} .
\label{Ham2}
\end{eqnarray}
As explained in Section~\ref{sec2.1}, to produce a TRP sweep in the detector
frame it is necessary to sweep $\dot{\phi}_{det}$ and $\dot{\phi}_{rf}$ 
through a Larmor resonance frequency. We choose (somewhat arbitrarily) to 
sweep through the Larmor frequency $\omega_{2}$: 
\begin{eqnarray}
\dot{\phi}_{det} & = & \omega_{2} +\frac{2at}{\hbar} + \Delta \nonumber\\
\dot{\phi}_{rf} & = & \dot{\phi}_{det} - \dot{\phi}_{4} .
\label{twoqbtdefs}
\end{eqnarray}
Here $\phi_{4}(t) = (1/2)Bt^{4}$ is the twist profile for quartic TRP, and we 
have introduced a frequency shift parameter $\Delta$ whose value will be 
determined by the sweep parameter optimization procedure of
Section~\ref{sec2.3}. Inserting eqs.~(\ref{twoqbtdefs}) into eq.~(\ref{Ham2}),
and introducing $\delta\omega = \omega_{1}-\omega_{2}$ and $b_{i}=\hbar
\omega_{i}^{rf}/2$ ($i=1,2$), we find
\begin{eqnarray}
\frac{\mbox{\~{H}}_{2}(t)}{\hbar} & = & \left[ -
         \frac{(\delta\omega +\Delta )}{2} + \frac{at}{\hbar}\right]\sigma_{1z}
          -\frac{b_{1}}{\hbar}\left[\,\cos\phi_{4}\,\sigma_{1x}+\sin\phi_{4}\,
           \sigma_{1y}\,\right] \nonumber \\
  & & \hspace{0.3in} +\left[ - \frac{\Delta}{2}+\frac{at}{\hbar}\right]
         \sigma_{2z} -
       \frac{b_{2}}{\hbar}\left[\,\cos\phi_{4}\,\sigma_{2x}\sin\phi_{4}\,
          \sigma_{2y}\, \right] \nonumber \\
  & & \hspace{0.7in} -\frac{\pi}{2}\, J\,\sigma_{1z}\sigma_{2z} .
\label{Ham3}
\end{eqnarray}
We see that both qubits are acted on  by a quartic TRP sweep in the detector
frame. In keeping with our earlier choice of sweeping through the 
Larmor resonance of the second qubit, we use $b_{2}$ in the definitions of the 
dimensionless time $\tau$, inversion rate $\lambda$, and twist strength 
$\eta_{\scriptscriptstyle 4}$:
\begin{eqnarray}
\tau & = & \left(\frac{a}{b_{2}}\right)\, t \label{twoqbttdef}\\
\lambda & = & \frac{\hbar a}{\left( b_{2}\right)^{2}} \\
\eta_{\scriptscriptstyle 4} & = & \left(\frac{\hbar B}{a^{3}}\right)\,
        \left( b_{2}\right)^{2} .
\label{twoqbte4def}
\end{eqnarray}
Since $\mbox{\~{H}}_{2}(t)/\hbar$ has units of inverse-time, and $b_{2}/a$ has
units of time (eq.~(\ref{twoqbttdef})), multiplying eq.~(\ref{Ham3}) by
$b_{2}/a$ and using eqs.~(\ref{twoqbttdef})--(\ref{twoqbte4def}) gives the 
dimensionless two-qubit Hamiltonian $\mbox{\~{H}}_{2}(\tau )$:
\begin{eqnarray}
\mbox{\~{H}}_{2}(\tau ) & = & \left[ -\frac{(d_{1}+d_{2})}{2}+
                               \frac{\tau}{\lambda}\right]\,\sigma_{1z}
                       -\frac{d_{3}}{\lambda}\left[\,\cos\phi_{4}\,\sigma_{1x}
                        +\sin\phi_{4}\,\sigma_{1y}\,\right] \nonumber\\
  & & \hspace{0.275in} +\left[ -\frac{d_{2}}{2} + \frac{\tau}{\lambda}\right]\,
        \sigma_{2z}
        -\frac{1}{\lambda}\left[\,\cos\phi_{4}\,\sigma_{2x}+\sin\phi_{4}\,
          \sigma_{2y}\,\right] \nonumber\\
  & & \hspace{0.65in} -\frac{\pi}{2}\, d_{4}\,\sigma_{1z}\sigma_{2z} ,
\end{eqnarray}
where
\begin{eqnarray}
d_{1} & = & \left(\frac{\delta\omega}{a}\right)\, b_{2} \nonumber\\
d_{2} & = & \left(\frac{\Delta}{a}\right)\, b_{2} \nonumber\\
d_{3} & = & \frac{b_{1}}{b_{2}} \nonumber \\
d_{4} & = & \left(\frac{J}{a}\right)\, b_{2} .
\label{ddefs}
\end{eqnarray}
As noted in Section~\ref{sec2.2}, $\mbox{\~{H}}_{2}(\tau )$ has a degeneracy
in the resonance frequency of the energy level pairs ($E_{1}\leftrightarrow
E_{2}$) and ($E_{3}\leftrightarrow E_{4}$). To break this degeneracy we add 
the term
\begin{equation}
\Delta H = c_{4}\, |E_{4}(\tau )\rangle\langle E_{4}(\tau )|
\end{equation} 
to $\mbox{\~{H}}_{2}(\tau )$, where $|E_{4}(\tau )\rangle$ is the instantaneous
energy eigenstate of $\mbox{\~{H}}_{2}(\tau )$ with eigenvalue $E_{4}(\tau )$. 
Our final Hamiltonian is then
\begin{equation}
H_{2}(\tau ) = \mbox{\~{H}}_{2}(\tau ) + \Delta H
\end{equation}
which is the Hamiltonian given in eq.~(\ref{simHam}). We see that 
$H_{2}(\tau )$ depends on the TRP sweep parameters ($\lambda$, 
$\eta_{\scriptscriptstyle 4}$), as well as on the parameters ($d_{1},\ldots ,
d_{4}$) and $c_{4}$. From eq.~(\ref{ddefs}) we see that $d_{1}$, $d_{2}$,
$d_{3}$, and $d_{4}$ are the dimensionless versions of, respectively, the
Larmor frequency difference $\delta\omega = \omega_{1}-\omega_{2}$, the
frequency shift parameter $\Delta$, the ratio $b_{1}/b_{2}=\gamma_{1}/
\gamma_{2}$, and the Ising coupling constant $J$.

\end{document}